\documentstyle[graphics]{mn}

\newif\ifAMStwofonts

\def\solm{M$_{\odot}\,$}

\def\etal{{\it et al.\ }}
\def\eg{{\it e.g.\ }}
\def\ie{{\it i.e.\ }}

\ifoldfss
  \newcommand{\rmn}[1] {{\rm #1}}

  \ifCUPmtlplainloaded \else
    \NewTextAlphabet{textbfit} {cmbxti10} {}
    \NewTextAlphabet{textbfss} {cmssbx10} {}
    \NewMathAlphabet{mathbfit} {cmbxti10} {} 
    \NewMathAlphabet{mathbfss} {cmssbx10} {} 
  \fi
  \ifAMStwofonts
    \ifCUPmtlplainloaded \else
      \NewSymbolFont{upmath} {eurm10}
      \NewSymbolFont{AMSa} {msam10}
      \NewMathSymbol{\upi}     {0}{upmath}{19}
      \NewMathSymbol{\umu}     {0}{upmath}{16}
      \NewMathSymbol{\upartial}{0}{upmath}{40}
      \NewMathSymbol{\leqslant}{3}{AMSa}{36}
      \NewMathSymbol{\geqslant}{3}{AMSa}{3E}

      \let\leq=\leqslant 
       
    \fi
  \fi
\fi 
 
\ifnfssone
  \newmathalphabet{\mathit}
  \addtoversion{normal}{\mathit}{cmr}{m}{it}
  \addtoversion{bold}{\mathit}{cmr}{bx}{it}
  \newcommand{\rmn}[1] {\mathrm{#1}}

  \newmathalphabet{\mathbfit} 
  \addtoversion{normal}{\mathbfit}{cmr}{bx}{it}
  \addtoversion{bold}{\mathbfit}{cmr}{bx}{it}
  \newmathalphabet{\mathbfss} 
  \addtoversion{normal}{\mathbfss}{cmss}{bx}{n}
  \addtoversion{bold}{\mathbfss}{cmss}{bx}{n}
  \ifAMStwofonts
    \ifCUPmtlplainloaded \else
      \UseAMStwoboldmath
      \makeatletter
      \new@mathgroup\upmath@group
      \define@mathgroup\mv@normal\upmath@group{eur}{m}{n}
      \define@mathgroup\mv@bold\upmath@group{eur}{b}{n}
      \edef\UPM{\hexnumber\upmath@group}
      \new@mathgroup\amsa@group
      \define@mathgroup\mv@normal\amsa@group{msa}{m}{n}
      \define@mathgroup\mv@bold\amsa@group{msa}{m}{n}
      \edef\AMSa{\hexnumber\amsa@group}
      \makeatother
      \mathchardef\upi="0\UPM19
      \mathchardef\umu="0\UPM16
      \mathchardef\upartial="0\UPM40
      \mathchardef\leqslant="3\AMSa36
      \mathchardef\geqslant="3\AMSa3E

      \let\leq=\leqslant 

    \fi
  \fi
\fi 
 
\ifnfsstwo
  \newcommand{\rmn}[1] {\mathrm{#1}}

  \DeclareMathAlphabet{\mathbfit}{OT1}{cmr}{bx}{it}
  \SetMathAlphabet\mathbfit{bold}{OT1}{cmr}{bx}{it}
  \DeclareMathAlphabet{\mathbfss}{OT1}{cmss}{bx}{n}
  \SetMathAlphabet\mathbfss{bold}{OT1}{cmss}{bx}{n}
  \ifAMStwofonts
    \ifCUPmtlplainloaded \else
      \DeclareSymbolFont{UPM}{U}{eur}{m}{n}
      \SetSymbolFont{UPM}{bold}{U}{eur}{b}{n}
      \DeclareSymbolFont{AMSa}{U}{msa}{m}{n}
      \DeclareMathSymbol{\upi}{0}{UPM}{"19}
      \DeclareMathSymbol{\umu}{0}{UPM}{"16}
      \DeclareMathSymbol{\upartial}{0}{UPM}{"40}
      \DeclareMathSymbol{\leqslant}{3}{AMSa}{"36}
      \DeclareMathSymbol{\geqslant}{3}{AMSa}{"3E}

      \let\leq=\leqslant 

    \fi
  \fi
\fi 
 
\ifCUPmtlplainloaded \else
  \ifAMStwofonts \else 
    \def\upi{\pi}
    \def\umu{\mu}
    \def\upartial{\partial}
  \fi
\fi

\title{Orbits Supporting Bars within Bars}
\author[Witold Maciejewski and Linda S. Sparke]
{Witold Maciejewski$^{1,2}$ and Linda S. Sparke$^1$\\
$^1$Department of Astronomy, University of Wisconsin,
475 N. Charter St., Madison, WI 53706\\
$^2$Theoretical Physics, University of Oxford, Oxford, OX1 3NP,
{\tt witold@thphys.ox.ac.uk}}

\begin{document}

\maketitle

\begin{abstract}
High-resolution observations of the inner regions of barred
disk galaxies have revealed many asymmetrical, small-scale 
central features, some of which are best described as secondary 
bars. Because orbital time\-scales in the galaxy center are short, 
secondary bars are likely to be dynamically decoupled from the 
main kiloparsec-scale bars. Here, we show that regular
orbits exist in such doubly-barred potentials
and that they can support the bars in their motion.
We find orbits in which particles remain on {\it
loops}: closed curves which return to their original positions
after two bars have come back to the same relative orientation.
Stars trapped around stable loops
could form the building blocks for a long-lived, 
doubly-barred galaxy. Using the loop representation, we can
find which orbits support the bars in their motion, and
what are the constraints on the sizes and shapes of
self-consistent double bars. In particular, it appears that
a long-lived secondary bar may exist only when an Inner Lindblad
Resonance is present in the primary bar, and that it would not
extend beyond this resonance.
\end{abstract}

\section{Introduction}
Recent high-resolution imaging, from space and from the ground, shows
hitherto-unsuspected small-scale structures at the centers of many
galaxies. In a number of barred galaxies, isophotal twists have been
seen within the central few hundred parsecs.
Small central bars, and occasional triply barred systems, are found: 
see Friedli (1996) for a review.
The inner bars appear to be oriented randomly with respect to the larger
bars (Buta \& Crocker 1993), as expected if they are dynamically
distinct subsystems. High-resolution CO maps of some disk galaxies show 
inner bars of molecular gas (\eg Devereux \etal 1992, Benedict \etal 
1996), but the presence of inner bars in infrared images (see \eg Friedli 
\etal 1996; Jungwiert, Combes \& Axon 1997; Mulchaey, Regan \& Kundu
1997) suggests they 
contain old stars, and do not consist purely of young stars and gas. 
A high frequency of double bars is seen in a multicolor imaging survey 
of early-type barred galaxies recently completed 
by Erwin \& Sparke (1999). The survey, performed with the WIYN telescope, 
and supplemented with archival HST images, covers a reasonably complete 
sample of 38 nearby ($z<$2000 km/s), bright, barred S0 and Sa galaxies in 
the field. At least $\sim$20\% of these galaxies
appear to possess secondary bars. In this paper, we will call the
larger bar the {\it big, main, primary} or {\it outer} bar. The smaller
bar will be referred to as the {\it small, secondary} or {\it inner}
bar. Like Friedli (1996), we prefer to avoid the term `nuclear bar',
since a small bar on scales of tens or hundreds of parsecs can exist 
even in the absence of a main kiloparsec-scale bar.

Within about 100 pc of a galactic center, 
orbital times are at least an order of magnitude less than those at a 
few kiloparsecs; thus a dynamically decoupled inner bar is likely to 
rotate faster than the 
outer structure. {\it The entire pattern is then not steady in any
reference frame.} Such decoupled 
secondary bars have been seen to form in numerical
simulations involving gravitating particles together with dissipative `gas
clouds' (Friedli \& Martinet 1993, Heller \& Shlosman 1994, Combes
1994, Shaw \etal 1995).  
The only systematic study of the orbital dynamics in a double-bar is
that of Pfenniger \& Norman (1990); they explored a weakly dissipative
form of the equations of motion for a particle in one double-bar
potential, finding a limit cycle in the plane of the bars, and
spheroidal strange attractors.
Orbits in the doubly barred potential do not have a
conserved integral of motion, and in principle they might all be
chaotic, exploring large regions of phase space. If the orbits 
are mostly chaotic, it is unlikely that such a system could be {\it
self-consistent}, so that the average density of all the stars on their
orbits in the time-varying potential adds up exactly to what is
needed to give rise to the potential in which they move. 

How can potentials including two independently rotating bars maintain 
themselves as gravitating systems? What are the conditions under which 
a gravitationally self-consistent double-bar structure could exist? We 
approach these questions by considering particle orbits in models that 
include two rigid bars rotating
at two constant, incommensurable pattern speeds. Such models cannot 
be fully self-consistent; Louis \& Gerhard (1988) have shown that
two bars which make up a self-consistent system must distort slightly as
they rotate through each other. Nevertheless, we are looking for density 
distributions 
that are most likely to be close to a self-consistent model, i.e. for 
those which support orbital families capable of hosting  sets of particles
that together recreate the assumed time-dependent density distribution. 
In our investigation we use the concept of the {\it loop} introduced
by Maciejewski \& Sparke (1997).
We explore constraints on double-bars that could maintain a likely
self-consistent structure.

The concept of the loop is presented in \S2. In \S3, we review the methods 
that we used to find and to follow loops
numerically in a doubly barred potential. In \S4, we describe our models 
for the density distribution in doubly barred galaxies, constructed on
the basis of parameters derived from observations. Various examples of 
loop families are given in \S5. In \S6, we closely examine loops supporting
a likely self-consistent model of a double bar. Limitations imposed by 
orbital structure on self-consistent doubly barred systems, and 
their implications for gas flows, are discussed in \S7. We summarize
our results in \S8. Analytical techniques 
that allow us to approximate the loops supporting a doubly barred density 
distribution, and to model the gas streamlines, are presented in Appendix
A. In Appendix B, we derive the formulae for a potential of a prolate
Ferrers bar.

\section{The concept of the loop}
A gravitational potential consisting of two concentric bisymmetric 
bars that are independently rotating is not stationary in any reference 
frame. Nevertheless, if the bars rotate in the same plane around their 
centers with angular velocities $\Omega_B$, $\Omega_S$, the potential
pulsates with a frequency $\omega_p = 2(\Omega_S-\Omega_B)$  in the 
frame rotating with one of the bars (note that the factor 2 comes from 
bisymmetry). 

In the stationary potential of a single bar, particles on a closed 
periodic orbit move along it, always staying on the same curve. 
Stable periodic orbits form ``backbones'' of a steady potential:
nearby orbits are trapped around
them. Orbits in a double-barred galaxy
will generally not be closed in any uniformly rotating frame of reference,
since particles there undergo two forcing actions with 
non-commensurable frequencies. 
We want to extend the definition of an orbit, in order to find 
closed curves which can similarly serve as backbones of a non-steady, 
doubly barred system. We postulate that these are curves which, when 
populated with
particles moving in a doubly barred potential, will return to their
original positions every time the bars come back to the same relative 
orientation. 
We call these curves {\it loops} --- they are a generalization 
of the closed periodic orbit in a single bar (see Maciejewski \& Sparke
1997). A particle that begins its orbit from a position along a given
loop returns to another point on the same loop
after the bars have realigned. Loops
change their shape as the bars rotate through each other.
Particles trapped around loops that stay aligned with the bars in their
motion could build up a long-lived, self-consistent,
doubly-barred galaxy, in the same way 
as particles trapped near closed periodic orbits in a singly barred
potential.

Although here applied to two rigid bars rotating with constant pattern 
speeds, this method works for any potential pulsating with a constant 
period. In particular, the two bars do not have to be rigid, or to 
rotate with constant pattern speeds.

\section{Methods}

\subsection{Formal statement of the loop problem}
We consider a phase space $\mathcal{S}$,
whose points are the coordinates and momenta of particles at a given time.
Then we define the mapping $\mathcal{P}(\mathcal{S})$, which 
assigns to each point in $\mathcal{S}$ the coordinates of the
corresponding particle after integrating its motion for one
relative period $T_p=2\pi/\omega_p$ of the bars. This mapping
$\mathcal{P}(\mathcal{S})$ can be applied repeatedly, leading to
consecutive {\it iterated maps}. For each point $A$ of the
phase space $\mathcal{S}$, the $\mathcal{P}$$^n(A)$ is called the 
{\it $n^{\rm th}$ iterate} of this point. The set of all iterates of a 
given point is called {\it the orbit} of this point {\it under the 
mapping} $\mathcal{P}(\mathcal{S})$ (see \eg Lichtenberg \& Lieberman 
1992). Thus defined, the orbits divide the 
phase space into non-overlapping invariant sets: no point 
can be on two distinct orbits.

A Hamiltonian system
in $N$ dimensions with a periodically varying potential can be written
as an autonomous system (one with a time-independent potential) in $N+1$
dimensions (see \eg Lichtenberg \& Lieberman 1992, Louis \& Gerhard 1988).
If we look only at orbits in the symmetry plane in which the bars rotate,
our 2-D potential changing periodically with a fixed frequency is 
equivalent to a 3-D autonomous Hamiltonian system. Regular orbits in 
this 3-D system are those 
that behave as if they have 3 integrals of motion: they are confined 
to a 3-D hypersurface of the 6-D extended phase space of the autonomous 
system. Our iterative mapping
suppresses the dimension corresponding to time, so the 3-D hypersurface 
of a regular orbit is represented by a 2-D set of points. Thus if a point
$A$ lies on a regular orbit in the double-bar system, its orbit under
$\mathcal{P}(\mathcal{S})$ is a 2-dimensional hypersurface in the 4-D 
space. Chaotic orbits will occupy 3 or 4  dimensions.
If a point in the phase space $\mathcal{S}$ lies on an orbit which 
closes after $N$ times the relative period,
its orbit under the mapping $\mathcal{P}(\mathcal{S})$
will consist of a 0-dimensional set of $N$ points. 

What about a point $A\in\mathcal{S}$, whose
iterates form a 1-D curve? This point is not on a periodic
orbit (since these form a 0-dimensional set 
under $\mathcal{P}(\mathcal{S})$),
but it corresponds to a particle moving on a special kind of a regular 
orbit (those are only confined to 2-dimensional hypersurfaces).
Every time the two bars return
to the same relative position, the iterates $\mathcal{P}$$^n(A)$
always lie on this 1-D curve: we have {\it a loop} as defined in
\S2. A particle that starts from any position on the loop will 
in general return
to another place on the same loop every time the bars align.
Since no point can be on two distinct orbits, and since in general
the iterates of one point on the loop fill the entire loop, every point 
$A$({\bf x},{\bf v}) which lies on a loop contains all the information 
about this loop. 

\begin{figure}
\resizebox{9cm}{!}
{\includegraphics{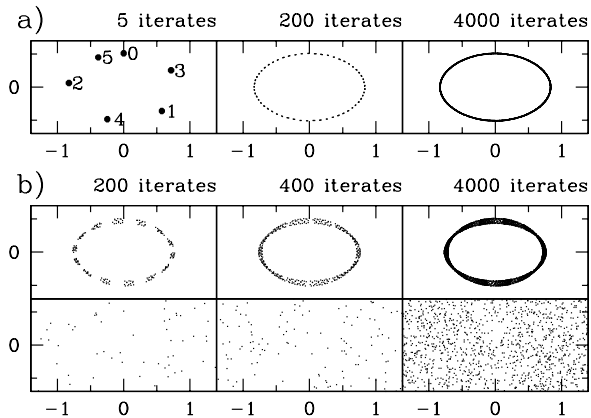}}
\vspace{-3cm}
\caption{{\it a)} A loop in position space, generated by iterates of one
of its points: 5 first iterates (left), 200 iterates (middle) and
4000 iterates (right). {\it b)} 200, 400 and 4000 iterates (left to 
right) in position space of a point that lies near a stable loop (top)
and one that does not (bottom). Points are plotted when both bars are 
horizontal. The corresponding diagrams in the velocity space have 
the same appearance.}
\label{makeloop}
\end{figure}

In Figure \ref{makeloop}a, we show how consecutive iterates of 
$A\in\mathcal{S}$ generate a loop in a doubly barred galaxy.
Figure \ref{makeloop}b shows 200, 400 and 4000 iterates $\mathcal{P}(\mathcal{S})$
for two points: the first lies near a stable 
loop, the second one does not.
For the near-loop solution, the points are confined 
to a ring around the stable loop. Iterates of a typical point 
far from the loop quickly disperse, covering much of
the phase space. Thus a stable loop can be found by
adjusting the coordinates of point $A\in\mathcal{S}$, until its orbit 
under $\mathcal{P}(\mathcal{S})$ converges onto a 1-D closed curve. 
Unstable loops can only be found in this way if we have a very good 
guess at the starting conditions. Our method requires following a 
single particle for many bar rotations, and any particle starting close
to an unstable loop will be lost by then. But since only stable loops 
can support the doubly barred systems, this is not a serious
disadvantage. Another way to look at stability is to compare the ring 
widths for
e.g. 100 and 200 iterations. They should be similar for a stable loop. 
For an unstable loop, a nearby particle moves away exponentially with 
time, increasing the width.

\subsection{Numerical methods of finding the loop}
\label{family}
To find iterates of point $A\in\mathcal{S}$, we
follow a particle in the $xy$ plane in a given potential 
$\Phi({\bf x},t)$, in which two bars
rotate independently about the $z$-axis.
We solve the equation of motion in a reference frame 
rotating with pattern speed $\Omega_B$, which is equal
to the angular speed of one of the bars. In the $xy$ plane 
of the barred disk, ${\bf \Omega_B} \perp {\bf r}$, and one can
write the equation of motion (see \eg formula 3-82 in Binney \& 
Tremaine 1987) as
\begin{equation}
\label{eqmot}
\ddot{\bf r} = -\nabla \Phi  - 2 ({\bf \Omega_B} \times \dot{\bf r}) 
+ |{\bf \Omega_B}|^2  \; {\bf r} .
\end{equation}
In Cartesian coordinates, it can be decomposed into its $x$ and $y$ parts:
\begin{eqnarray}
\label{eomx}
\ddot{x} & = & -\frac{\partial \Phi}{\partial x} 
+ 2 \Omega_B \dot{y} + \Omega_B^2 x , \\
\label{eomy}
\ddot{y} & = & -\frac{\partial \Phi}{\partial y} 
- 2 \Omega_B \dot{x} + \Omega_B^2 y .
\end{eqnarray}
We integrated the equations above using a variable-order, 
variable-step Adams integrating subroutine for first order 
differential equations. This predictor-corrector method is well
suited to our problem, for which we expect multiply periodic solutions.
Under the forcing action of two bars, a particle can be 
given a kinetic energy large enough to escape. 
Any particle departing from the potential center to distances
a few times larger than the extent of the outer bar was
considered to be lost, and we looked only for loops in and around the
bars.

We expect symmetric loops to be most 
important in symmetric bars. Therefore we restricted our search to 
the loops which are symmetric with respect to both $x$- and $y$-axes
when the bars are aligned on the $x$-axis: we reflected the successive
iterates $\mathcal{P}$$^n(A)$ about the two axes
so that they occupy the $0\leq \varphi < \pi/2$ region only.
For symmetric loops the $y$-velocity on the $y$-axis is zero,
and one can explore various initial $x$-velocities $v_x$
for a given $y$-axis crossing coordinate $y$. To help us find the 
right $v_x$, we used the epicyclic approximation of Appendix A
(see also Maciejewski \& Sparke 1997). 
In general, consecutive iterates  occupy the interior
of a thick ring, but there is a range of $v_x$ for which the ring
generated by a particle starting at that point looks almost 
1-dimensional. We assume that the real loop is nearby, and we need a 
practical definition of the loop: given $N$ ({\bf x},{\bf v}) pairs, 
we want to know if they lie on a 1-D curve.
In principle this is not possible: there is a curve through any finite set
of points. Nevertheless, we are looking for simple curves. 
The ideal choice might be to find the shortest line connecting 
the phase space positions of all the points,
and to minimize its length as $v_x$ varies. It is known though,
that this task (the `traveling salesman problem') is 
very expensive numerically: the number of operations needed to solve it 
grows exponentially with the number of points. Instead, we search for
the loop by varying $v_x$ in order to minimize the width of the ring 
in the $xy$ plane formed by the iterates of the assumed starting point. 

\begin{figure}
\resizebox{9cm}{!}
{\includegraphics{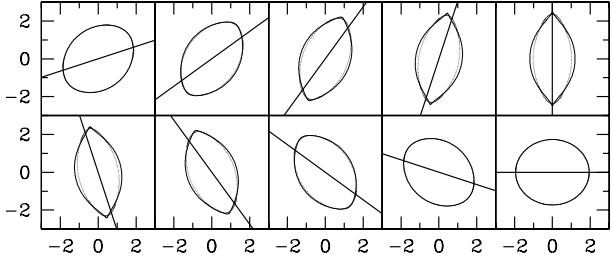}}
\vspace{-5cm}
\caption{Evolution of a loop in Model 1 in the frame rotating with 
the main bar, which remains horizontal. The straight solid line marks 
the major axis of the small bar. The epicyclic approximation is plotted 
as a solid line, the true loop is dotted. Units are in kiloparsec.}
\label{compareloop}
\end{figure}

Before minimizing, 
we adjusted each loop candidate for flattening in the $xy$ plane
by finding the points with maximum $x$-coordinates 
$x_{\rm max}$ and maximum $y$-coordinates $y_{\rm max}$. Then we rescaled
the $y$-coordinates, multiplying them by $y_{\rm max}/x_{\rm max}$,
to give a roughly circular ring. This was divided into 20 sectors, and 
the points with the smallest and largest radii were found in each sector.
The difference of these extremal radii gives the width of the ring in 
each of the 20 sectors. We define the overall ring width $w(y,v_x)$
as the maximum width, or its average over the sectors. Now, we can 
minimize $w(y,v_x)$ with respect to the $x$-velocity and get the
minimum width $w(y)$. Both 
minimizations proved to be useful, and the method itself is efficient and 
fast. In principle, we should be able to bring the ring width 
arbitrarily close to zero by increasing the number of iterates and
sectors, and adjusting the initial velocity $v_x$. In practice,
20 sectors and 400 iterates proved to be good and efficient
enough. A clear minimum 
is present at the same location for both estimates. We assume that
the loop is generated by a particle with the starting $x$-velocity
for which the width is minimal. The evolution of one loop thus defined 
is shown in Figure \ref{compareloop}, together with the corresponding solution of the
epicyclic approximation for comparison. The epicyclic approximation 
remains close to the real loop, which justifies using this 
approximation to begin the loop search. For non-symmetric loops,
we would have to minimize with respect to 2 parameters; for example 
the initial values of $v_x$ and $v_y$.

Iterates of a given point are not always spread uniformly around the
ring, but sometimes collect into several distinct clumps. This happens 
when the frequency
$\omega$ with which the particle circumnavigates the loop is almost
commensurable with the potential oscillation frequency $\omega_p$.
If these frequencies remain close to a lower order resonance, the number 
of clumps is small, and there are few points outside the
clumps. This may produce an additional spurious minimum in the average 
ring width, but the minimum is narrow and changes quickly with the
starting $y$-coordinate when the whole loop family 
is examined (see below), so its character is easily recognized.

In a single bar, some of the periodic orbits are self-intersecting; we
expect to find self-intersecting loops in double bars as well. Our method
of minimizing the ring width only in position- or velocity-space obviously 
breaks down there. We could consider the loop in the 4-dimensional phase
space, where it never intersects itself, at the cost of a considerably
larger computational expense and less clear visual examination of the
results. Alternatively, we can 
calculate the width for part of the ring only. For example, if the loop
intersects itself on the $x$-axis, these regions can be excluded from
the calculation of the width. This procedure proved to be 
very fast and robust, and has been used exclusively.

The epicyclic approximation provides at most one solution for the loops
and orbits at a given guiding radius. Additional loops, not predicted 
by the epicyclic approximation, can be present and may be dynamically 
important in the system of two bars. The only way to find them is to 
search the phase space of the initial conditions $(y,v_x)$ for the ring 
width $w(y,v_x)$ values. In the following sections, we call this
procedure the phase plane search.
Sets of points $(y,v_x)$ with low $w$ that group along continuous
lines should indicate loop families: continuous functions $v_x(y)$ for 
a range of $y$ and $v_x$. The larger the region of low $w$ around the 
line marking a loop family, the more orbits such a loop family traps 
around itself, and thus the more useful `backbone' for building the 
galaxy it is.

We searched for loop 
families by changing the initial value $y$ on the $y$-axis, and then
adjusting the velocity $v_x$, so that the new pair $(y,v_x)$ generates 
another loop. By analogy with periodic orbits in a single bar,
we call $v_x(y)$ {\it the characteristic curve} for a given
family. In our search we noticed that
the minimal width of the ring of iterates can sometimes be large, but
the loop family preserves continuity. We therefore accepted some fairly
shallow minima in the automatic loop search. Doing so could 
derail the search from one loop family to another, if they come
close to one another in the phase space $(y,v_x)$. To clarify when the 
border between different families has been crossed, we searched for
orbital families twice. Once we did it by following particles starting
on the $y$-axis, when the bars were both aligned on the $x$-axis, 
as described above. We did a second search, starting 
the particle on the $y$-axis when the big bar was still on the $x$
axis, but the small bar was perpendicular to it.
If the search algorithm loses the loop, one can go back to the phase plane
display: identifying the region where the loop was lost often helps
to answer why it was lost. Examples of this procedure will be given
in sections 5 and 6.

\section{Models based on observations}
\subsection{The structure of the models}
\label{modpar}
To minimize the number of free parameters characterizing our
models, we followed Athanassoula (1992a) in adopting analytical
forms for the various potential components. Our potential consists of a 
bulge, a disk and two bars. There is no halo: in these 
2-dimensional calculations its contribution to the rotation curve
is included in the disk potential. Mass can be distributed
arbitrarily into spherical and disk-like components, as long as 
they together make a realistic rotation curve.

\begin{table*}
\begin{minipage}{80mm}
\caption{12 parameters for our double-barred galaxy models}
\begin{tabular}{lrr} \\
[-.4cm]
                                         &   Model 1      & Model 2 \\
\hline\\[-.3cm]
\multicolumn{3}{c}{\it disk parameters}    \\[.3cm]
$R_{\rm max}=\sqrt{2}R_0$                & $20 \ \rm{kpc}$         & $20 \ \rm{kpc}$ \\
$v_{\rm max}=2\sqrt{\pi G R_0 \sigma_0}/3^{0.75}$ & $164 \ \rm{km\;s}^{-1}$       & $164 \ \rm{km\;s}^{-1}$ \\
\multicolumn{3}{c}{\it bulge and bar parameters} \\[.1cm]
total mass within 10 kpc $M_{\rm tot}$   & $5\times10^{10}$ \solm 
                                         & $5\times10^{10}$ \solm \\
central density $\rho_b+\rho_{0B}+\rho_{0S}$&$1\times10^{10}$ \solm$\rm{kpc}^{-3}$ 
                                         & $4.8\times10^{10}$ \solm$\rm{kpc}^{-3}$ \\
primary bar semimajor axis $a_B$         & $7 \ \rm{kpc}$          & $6 \ \rm{kpc}$ \\
primary bar quadrupole moment  $Q_m$     & $2.25\times10^{10}$ \solm $\rm{kpc}^2$ 
                                         & $4.5\times10^{10}$ \solm $\rm{kpc}^2$ \\
Lagrangian radius of the big bar & $1.2 a_B$      & $1.0 a_B$ \\
axial ratio of the primary bar $a_B/b_B$ & $2.5$            & $2.5$ \\
axial ratio of the secondary bar $a_S/b_S$ & $2.5$          & $2.0$ \\
ratio of bar semimajor axes $a_S/a_B$  & $0.6$            & $0.2$ \\
ratio of bar masses $M_S/M_B$          & $0.6$            & $0.15$ \\
pattern speed of the secondary bar $\Omega_S$ &$42 \ \rm{km\;s}^{-1}\rm{kpc}^{-1}$& $110 \ \rm{km\;s}^{-1}\rm{kpc}^{-1}$ \\
\hline
\end{tabular}
\end{minipage}
\end{table*}

The volume density of the bulge is given by a modified Hubble profile
\begin{equation}
\rho(r) = \rho_b (1+r^2/r_b^2)^{-1.5} .
\end{equation}
It requires two free parameters: the bulge central density $\rho_b$ and
the characteristic radius $r_b$. The surface density of the disk is 
defined by a Kuzmin-Toomre profile (Kuzmin 1956; Toomre 1963)
\begin{equation}
\sigma(R) = \sigma_0 (1+R^2/R_0^2)^{-1.5} ;
\end{equation}
like the bulge, this requires two parameters: $\sigma_0$ and $R_0$.
Following Athanassoula, we adopted the Ferrers ellipsoid to describe both
bars, with volume density 
\begin{equation}
\label{barden}
\rho(x,y,z) = \left\{ \begin{array}{ll}
                       \rho_0(1-m^2)^n & \mbox{if $m<1$} \\
                       0               & \mbox{otherwise ,}
                       \end{array}
              \right.
\end{equation}
where in Cartesian coordinates $m^2 = x^2/a^2 + y^2/b^2 + z^2/c^2$.
Thus the isodensity surfaces of our bars are concentric ellipsoids 
of constant axis ratio, and each bar is described by 5 free parameters: 
the central density 
$\rho_0$, the Ferrers exponent $n$, and 3 semi-axes. We adopt prolate bars,
and will set $n=2$, which reduces the number of free parameters to 3. 
A larger index $n$ allows the higher derivatives of the potential 
to be continuous, which in turn reduces the fraction of chaotic orbits 
in the solution. If there are loops supporting doubly barred
potentials, they are more likely to be found for bars with higher $n$.
In Appendix B, we show how the potential of a prolate Ferrers bar can be 
obtained by reducing Pfenniger's (1984) solution for a triaxial bar. 
Both bars rotate about the $z$-axis with respect to the inertial 
frame; their pattern 
speed, or equivalently the Lagrangian radius, constitutes
a 4th free parameter for each bar in our case. Thus the whole system
is modeled with 12 parameters.

\begin{figure}
\resizebox{9cm}{!}
{\includegraphics{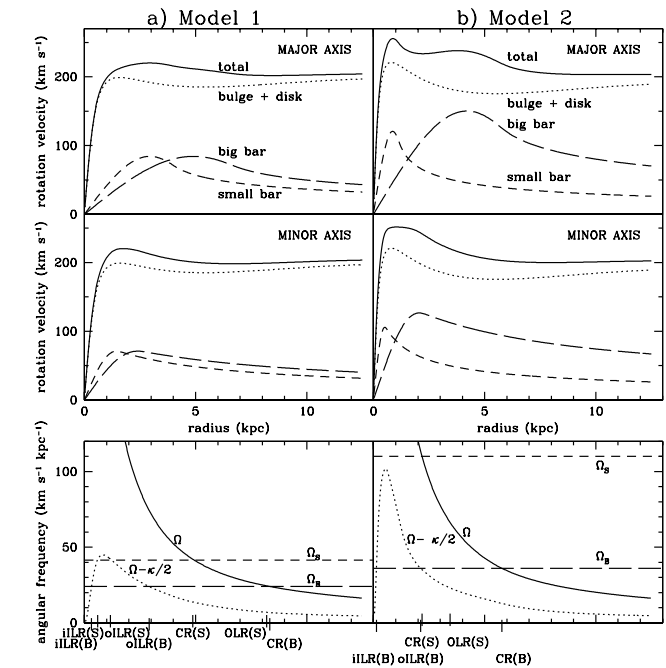}}
\caption{The rotation curve ($\sqrt{R \; d\Phi/dR}$) and its components 
along the major axis (top panels) and minor axis (central panels) for 
{\it a)} Model 1 and {\it b)} Model 2. The bottom panels show the angular 
frequency curves in the azimuthally averaged potential that determine the 
positions of resonances in the big bar
(B) and small bar (S). We marked the inner and outer ILRs (iILR
and oILR respectively), the corotation (CR), and the Outer
Lindblad Resonance (OLR).}
\label{rotcurve}
\end{figure}

The parameters for the two models considered in this paper are given 
in Table 1.
The potentials of the bulge, disk, and the big bar are chosen on the 
basis of Athanassoula's (1992a) `standard model', constructed by Wozniak
(1991) from a set of potentials inferred from observations of 
16 SB0 galaxies and one SBa. The disk parameters are the same as in 
Athanassoula's model.
The parameters of the secondary bars were chosen on the
basis of the observations listed in the next subsection. In Figure 
\ref{rotcurve}, we show the rotation curve, calculated from forces 
along major and minor axis of the bar at bars aligned, and the curve 
of $\Omega-\kappa/2$ in the azimuthally averaged potential for both models.

Model 1 was designed to explore how far the secondary bar can extend in
a doubly barred potential. It is the same model as the one used by 
Maciejewski \& Sparke (1997). The parameters of the axisymmetric part 
match Athanassoula's average observed values almost exactly. But
the main bar is weaker than the observed average, although falling 
within the range found by Wozniak (1991). 

In Model 2, the quadruple moment of the primary bar, its Lagrangian 
radius, and its axial ratio are the same as in Athanassoula's model. Since
we use $n$=2 Ferrers bars, whose density drops faster with radius
than in the $n$=1 bar modeled by Athanassoula, we extended the 
main bar all the way to its Lagrangian radius at 6 kpc, so the mass 
distributions are roughly
similar. We increased the total central density to a value twice than 
that in Athanassoula's model, so the secondary bar comes close to having 
an ILR (Fig. \ref{rotcurve}). Of the 16 galaxies in the Athanassoula \& Wozniak sample 
(Wozniak 1991), only two have such a big maximum of the $\Omega-\kappa/2$
curve, but the other parameters for the
axisymmetric part fit these observations well. Both bars
are stronger here than in Model 1, though still weaker than Wozniak's
observed average.

\subsection{Observed parameters of doubly barred galaxies}
Friedli (1996) reviews available observations of possible double bars
and estimates their basic parameters. We follow his notation here, in 
which three parameters are defined:
\begin{enumerate}
\item the maximum ellipticity in the barred region 
$e^{\rm max}=\max_a|1-b/a|$, where $a$ and $b$ are the semimajor and
semiminor axes of the fitted ellipses (note that this is the maximum 
ellipticity of the total light distribution, not of the bar only),
\item the length ratio of the bars $\beta = l_B/l_S$, where $l_B$, $l_S$
are the semi-major axes of the isophote ellipse fits at maximum or minimum 
ellipticity that corresponds to the main and secondary bar, respectively,
\item the luminosity ratio of the bars $\gamma = L_B/L_S$: the ratio of 
total luminosities within the isophote of maximum (minimum) ellipticity 
(note that thus defined, the inner bar's luminosity is strongly influenced 
by the light from the bulge).
\end{enumerate}
Wozniak \etal (1995) discuss how the last two parameters above depend on whether
they are derived at minimum or maximum ellipticity. Mulchaey, Regan \&
Kundu (1997) present another, smaller but rather unbiased sample of
doubly barred galaxies seen in K-band images.
Average values at minimum ellipticity of parameters defined above
for both samples (13 and 6 doubly barred galaxies respectively)
are given in Table 2. There is an observational bias towards lower 
values of $\beta$ and $\gamma$, because such systems are most easily 
found. Also $ e^{\rm max}$ for the small bar may be underestimated 
because of finite spatial resolution.

\subsection{Model parameters versus observational constraints}
To make a crude comparison between these models and the observed
double bars, Erwin (1998, private communication) performed ellipse 
fitting on both our models in face-on position for a
relative bar orientation of 60\degr,
and derived the values of $\beta$, $\gamma$, $e^{\rm max}_S$ and 
$e^{\rm max}_B$. Instead of surface brightness, the surface density 
was used. Since the contribution of the dark halo in our models is
included in the disk mass distribution, the halo appears luminous
here. It is not important though, since we are interested in the
central parts of the galaxy, where the luminous matter is dominant.
We calculate the luminosity of each bar as
an integral over the interior of the ellipse with maximum
or minimum ellipticity. We consider the parameters derived at the
maximum ellipticity (max-$e$) to be better defined, because in real 
galaxies the radius of minimum ellipticity (min-$e$) at the main bar 
depends on 
the inclination angle. Table 3 gives the derived parameter values.

\begin{table}
\caption{Observed parameters of doubly barred galaxies (Friedli 1996 / Mulchaey \etal 1997)}
\begin{tabular}{lrrr} \\
\hline
                 & average & min & max \\
\hline
$\beta $           & 7.2/4.9     & 3.7/1.83 & 18.0/8.8 \\
$\gamma $          & 4.0/-       & 2.0/-    &  7.5/-   \\
$ e^{\rm max}_S$   & 0.34/0.30   & -/0.2    & -/0.43  \\
$ e^{\rm max}_B$   & 0.55/0.41   & -/0.2    & -/0.6   \\
\hline
\end{tabular}
\end{table}
The $\beta$ and $\gamma$ parameters for Model 1
derived with both `min-$e$' and `max-$e$' methods are too small
compared to the observed range (Table 2). Thus the secondary bar in
Model 1 is too large; despite that, we explore this model to check if such
a configuration is theoretically possible. 

In Model 2, we see well defined ellipticity maxima, with a minimum 
between them. Note that both `max-$e$' and `min-$e$' methods give 
bar sizes smaller than the major axis of the Ferrers bar: in Model 2
the primary bar's
half-length is estimated to be 3.58 and 5.58 kpc respectively, when 
the semi-major axis of the Ferrers primary bar is 6.0 kpc.
The measured values of the half-length of the secondary bar are 0.62
and 0.92 kpc respectively, when we set $a_S$=1.2 kpc. 
The $\beta$ and $\gamma$ parameters in `min-$e$' and 
`max-$e$' estimations have very similar values. 
Note that the mass in the inner parts of the galaxy is dominated by the 
bulge, and this makes $\gamma$ much smaller than the mass ratio of 
the bars. For Model 2, the $\beta$ and $\gamma$ values derived by the 
ellipse-fitting method correspond very well to
the parameters in the observed samples (Table 2). 

The only parameter of Model 2 that does not conform to
the observed range, is the `measured' maximum ellipticity 
of the secondary bar. This is only half as large as observed 
average, because the central parts of the model are dominated by 
our strong spherical bulge.
It is very likely then, that distinct secondary bars are best
observed against weak bulges, which cannot induce an ILR in them,
unless the bars themselves are highly centrally concentrated.

From Figure \ref{rotcurve} we see that the corotation of the secondary bar in Model 2 
($a_S$=1.2 kpc) is at 2.3 kpc, well outside the bar.
Such a setup was forced after Model 1 failed to provide orbits supporting
outer parts of the secondary bar --- see the following section.
The slowly rotating secondary bar seems
to be consistent with the most successful numerical model by Friedli \& 
Martinet (1993, Set II). They observe a secondary bar of 2.5 kpc size 
with corotation at 3.5 kpc, which makes $r_L/a \approx$ 1.4, compared to
1.9 in our case. Thus the slowly rotating secondary bar appears to
be justified by numerical models.

\begin{table}
\caption{Model parameters derived from isophote fitting}
\begin{tabular}{lrrrr} \\
\hline
                 & \multicolumn{2}{c}{Model 1} & \multicolumn{2}{c}{Model 2} \\
\hline
                 & min-$e$ & max-$e$          & min-$e$ & max-$e$ \\
\hline
$\beta$          & 3.07    & 1.98             & 6.08    & 5.79 \\
$\gamma$         & 2.38    & 1.65             & 3.55    & 3.58 \\
$e^{\rm max}_S$ & \multicolumn{2}{c}{0.118}   & \multicolumn{2}{c}{0.159}\\
$e^{\rm max}_B$ & \multicolumn{2}{c}{0.182}   & \multicolumn{2}{c}{0.428}\\
\hline
\end{tabular}
\end{table}

\section{Examples of loop search methods}
\subsection{Loop families in a single bar}
To illustrate the methods which allow us to find loops in doubly 
barred galaxies, we show how the loop search is performed for a 
single bar. Here, we can choose arbitrarily the period of integration
(we chose it to be very small)
used to generate consecutive iterates of a given point. If this point 
lies on a periodic orbit, all its iterates will also lie on this orbit:
in a single bar our loop finding technique locates the closed 
periodic orbits. If the orbit is stable,
nearby particles should generate points populating a ring around 
this orbit.

\begin{figure}
\vspace{-2cm}
\resizebox{8cm}{!}
{\includegraphics{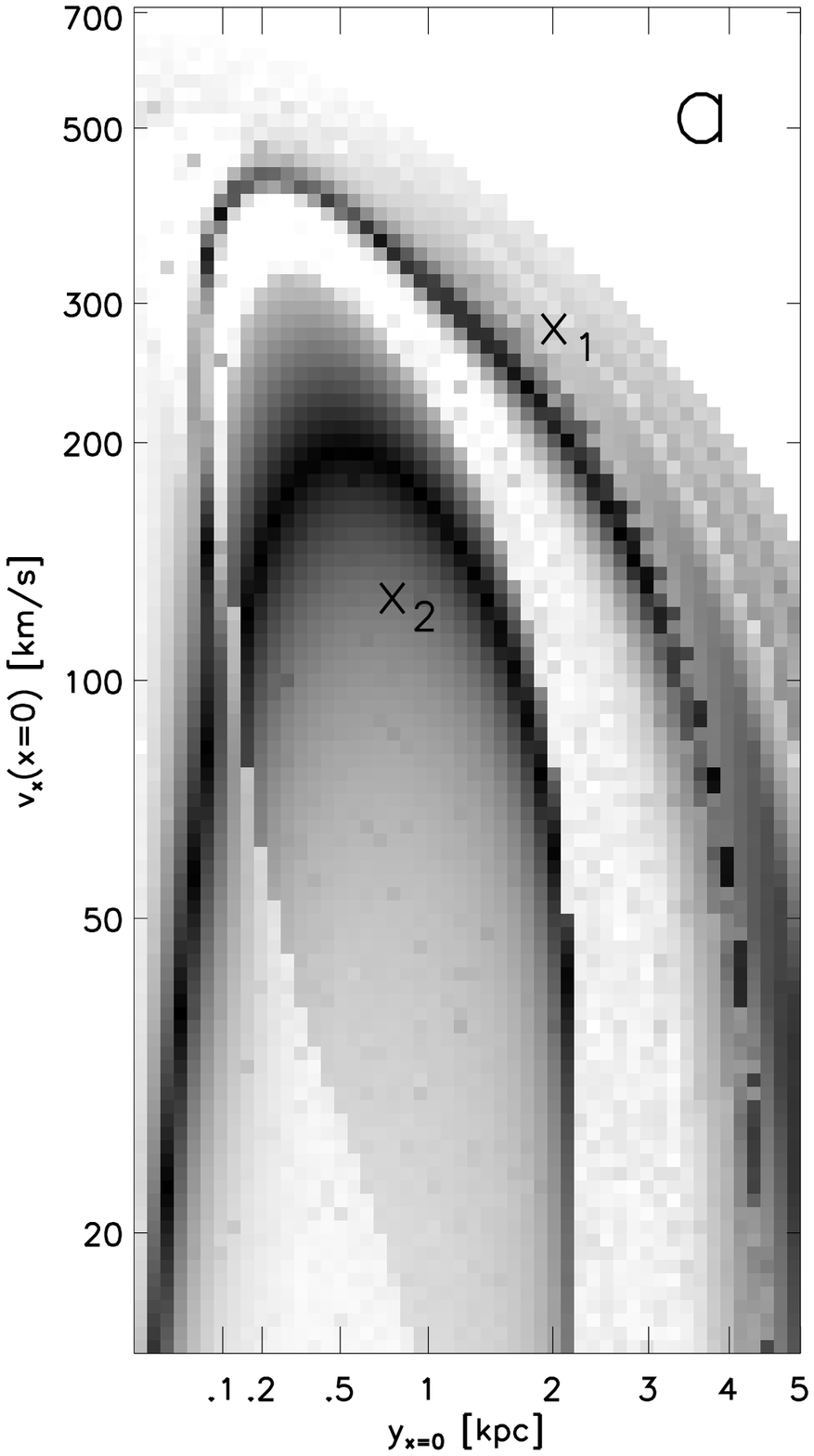}}
\resizebox{9cm}{!}
{\includegraphics{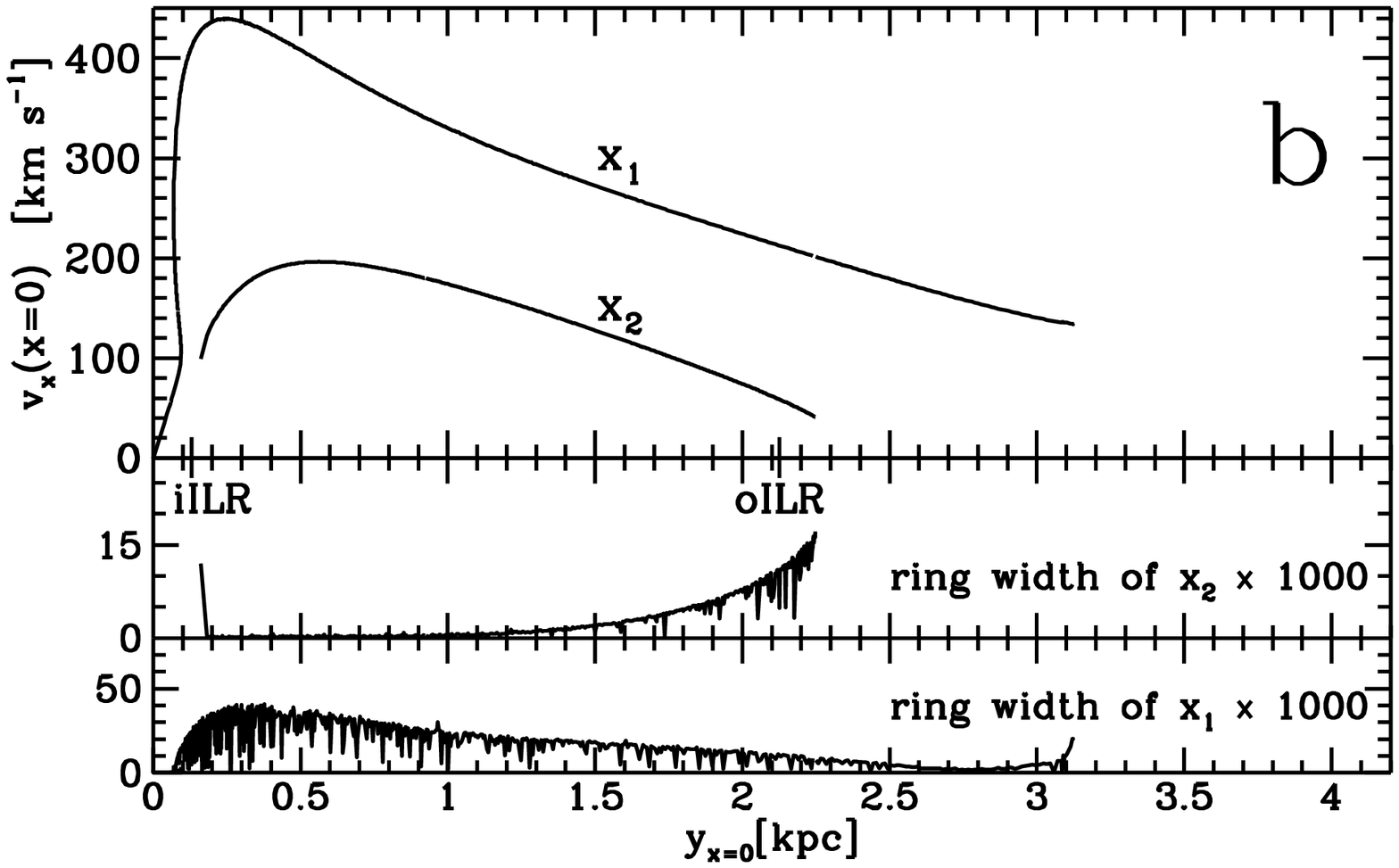}}
\vspace{-3cm}
\caption{{\it a)} Phase plane search for particles in Model 2 without
the secondary bar, which start on the minor axis of the bar.
The colour reflects how strongly the first 400 iterates of a particle with
given starting values ($y,v_x$) are confined to a 1-D curve.
Darkest areas mark the near-loop solutions (smallest ring widths $w$).
{\it b)} The characteristic curves for Model 2 without the secondary bar.
The diagrams of minimal ring widths $w(y)$ along each family are attached.}
\label{m2sgl}
\end{figure}

We performed the search for Model 2, with the mass of the inner bar 
redistributed into the axisymmetric bulge, so that the central density
and the total mass defined in Table 1 remain unchanged.
Figure \ref{m2sgl}a illustrates the phase plane search defined in section 3.2. 
Two dark arches surround the main stable orbital families; 
in the notation of Contopoulos \& 
Papayannopoulos (1980), these are the $x_1$ family, elongated parallel 
to the bar, and the $x_2$ family, oriented perpendicular to the bar. 
Thick dark arches,
which overwhelmingly dominate the phase plane, mean that 
many quasi-periodic orbits are trapped around these periodic orbits.
The unstable $x_3$ family is absent from this plot, because particles
are not trapped around it.

The characteristic curves $v_x(y)$ for the $x_1$ and $x_2$ families
are shown in Figure \ref{m2sgl}b, 
together with the minimum values of the ring width $w$ defined in 
section 3.2, describing how close a set of iterates 
of the starting point given by $(y,v_x)$ falls to a 1-D curve.
The smooth appearance of the widths 
assures us that we follow a single loop family. In particular,
when the orbital family $x_2$ disappears at large $y$,
the ring widths increase rapidly.
In \S6, we will compare these orbits to the loops in the doubly barred
Model 2, which is close to being self-consistent.

\subsection{Loop families in a generic doubly barred potential: Model 1}
Here we discuss the loop families in Model 1, the first of our doubly
barred potentials. It is more general than Model 2, because the bars are not 
in mutual resonance, and the secondary bar is relatively large.
Both bars are rapidly rotating: they extend to $\sim 85 \%$ of their 
Lagrangian radii. To enable a comparison of the calculated loops
to those from an epicyclic approximation, we made
the bars weaker than in Athanassoula's (1992a) standard model. Both
bars have two ILRs: those of the main bar are at 0.32 kpc and 2.9 kpc,
and of the small bar at 0.6 kpc and 1.3 kpc.

\begin{figure}
\vspace{-2cm}
\resizebox{8cm}{!}
{\includegraphics{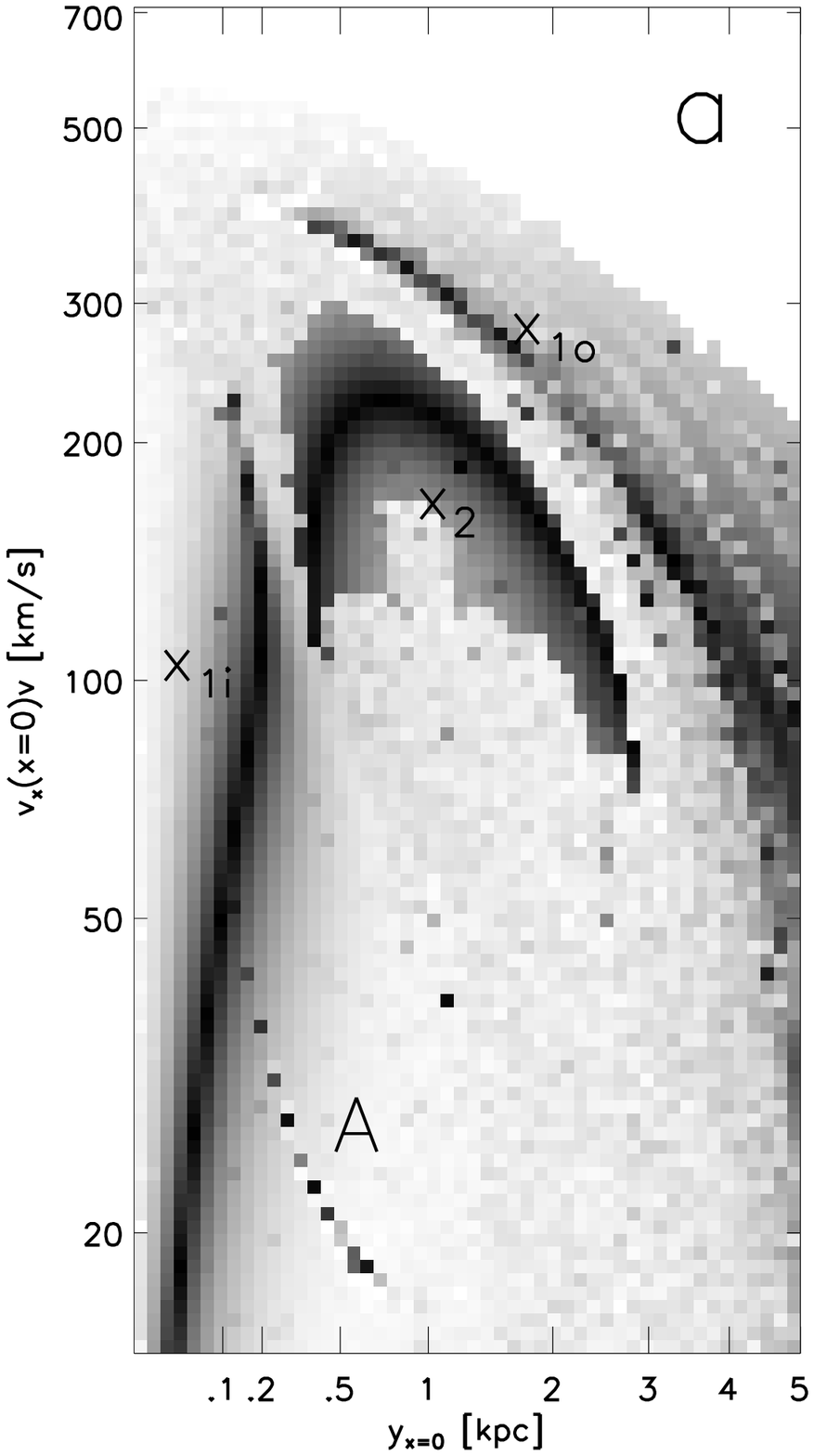}}
\resizebox{9cm}{!}
{\includegraphics{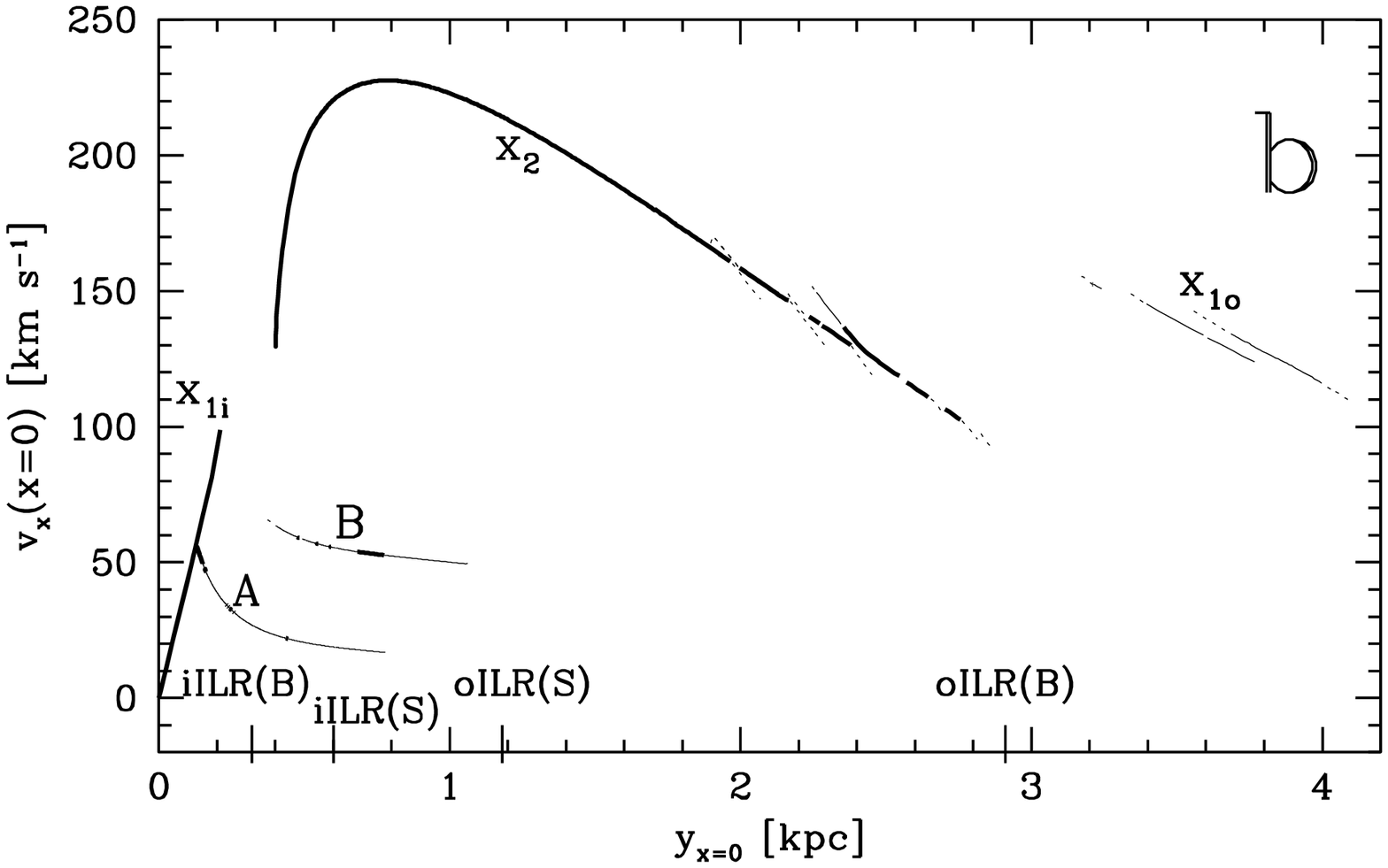}}
\vspace{-3cm}
\caption{{\it a)} As Fig.\ref{m2sgl}a, but for particles in Model 1, 
which start on the minor axis of the bar when the bars are aligned.
{\it b)} The characteristic curves for Model 1. Loop families and
resonances are labeled: $x_{\rmn{1i}}$ and $x_{\rmn{1o}}$ are the
inner and outer parts of the $x_1$ family, as described in the text. 
The thickness of the
line corresponds to the ring width: the loops with clear minima
in width are represented by the thickest lines, the dotted lines
mark areas where loops are no longer well defined. The labels $A$ 
and $B$ refer to loop families described in section 5.2.3}
\label{m1}
\end{figure}
 
Figure \ref{m1}a presents a phase 
plane search for loops which are symmetric with respect to both
$x$- and $y$-axes at the time when the bars are aligned on the $x$-axis.
As in Figure \ref{m2sgl}a,
two dark arches surround the stable loop families,
though the top arch is broken in the middle. Extending the 
notation of orbital families, we call the loop family that 
corresponds to the $x_2$ orbital family of the main bar,
the $x_2$ loop family. It appears clearly in Figure \ref{m1}a, along with a 
wide dark stripe of near-loop solutions, which shows that particles can 
easily be trapped around this family. The $x_1$ loop family corresponds
to the $x_1$ orbits in the outer bar. We see the inner part of the 
$x_1$ loop family clearly out to the radius where the
bifurcation giving rise to the $x_2$ family takes place. The family
abruptly discontinues above this point ($v_x\approx$250 ---
when we refer to positions in $v_x(y)$ diagrams in this and the 
following sections, the length units are kpc, and the velocity
units are km s$^{-1}$). The inner part of the $x_1$ loop family remains 
inside the inner ILR of the main bar.
The $x_1$ loop family can be detected again for $0.4<y<2.0$, between 
the two ILRs of the main bar. But here it is not accompanied by a 
wide dark region of near-loop solutions, indicating that
it may be not able to trap many particles around itself. Our plane search
finds no clear $x_1$ loop family in the region $2<y<3$, close to the 
outer ILR of the main bar; there may be
no loops there, or the resolution of our search may be too low --- it
shows only a grey region of near-loop solutions in this $y$ range. For 
larger $y$ this family is seen again, together with the accompanying
dark grey band of near-loop solutions. The discontinuity of the $x_1$
family at large $v_x$ seems to be common in our double-bar potentials;
therefore we
divide this family into an inner part $x_{\rmn{1i}}$ inside the $v_x(y)$ 
maximum and an outer part $x_{\rmn{1o}}$ outside it. 

In addition to the
$x_1$ and $x_2$ loop families, there is an interesting dark line marked 
in Figure \ref{m1}a by $A$, which departs from the inner $x_1$ family in
the lower left corner of the phase plane ($y\approx 0.13$, 
$v_x\approx 60$). It is examined separately below. No other 
loop families can be clearly seen in Figure \ref{m1}a. 
In particular, there seem to be no stable loops corresponding to $x_2$ 
orbits of the small bar, although this bar has a pretty wide zone
between inner and outer ILRs.

All the loop families that we have found for Model 1 are displayed 
in Figure \ref{m1}b; they are described in detail in the rest of this section.
The thick lines correspond to well defined loops of small ring width;
thin lines represent loops that are still well defined as single minima,
but with considerable ring widths. The broken lines correspond 
to regions where loops were not well defined, mostly because at fixed
$y$ the ring width had multiple minima with $v_x$, or the minimum value
of $w$ was rapidly growing with $y$.

\subsubsection{The $x_2$ loop family}
We were able to follow the $x_2$ family as a continuous minimum of the 
ring width from ($y$=0.4, $v_x$=130) to ($y$=2.2, $v_x$=140), which 
remains inside the zone between inner and outer ILRs
of the main bar. The $x_2$ loops still can be found in the range 
2.2$<y<$2.8, but the $x_2$ family is broken in this region.
Our attempts to follow it fail, because multiple
ring width minima compete with one another as $y$ changes.

A representative set of the $x_2$ loops 
in the frame rotating with the primary bar is shown in Figure \ref{m1loops}a. Here 
we confirm the hypothesis that we developed on the basis of the epicyclic 
approximation (Maciejewski \& Sparke 1997): 
the outermost $x_2$ loops correspond to the $x_2$
orbits of the main bar and remain roughly perpendicular to it, while the
inner loops of this family follow the secondary bar in their motion, 
and correspond to the $x_1$ orbits of the secondary bar. However, 
we cannot trace the $x_2$ loops all the way to the center, while
the $x_1$ orbits of a single bar generally do extend to the center. 

Throughout this paper we assume that our 2 bars are rigid and uniformly 
rotating. But the structure of the $x_2$ loops in Figure \ref{m1loops} tells us that 
it is a crude approximation only:
the secondary bar, as outlined by its orbital (loop) structure, pulsates
and accelerates as it moves through the main bar. The loops supporting 
the small bar are much thicker when it is parallel to the main bar: the 
minor axes of the inner loops are up to 1.8 times as large as when the 
bars are perpendicular.
So if a secondary bar is self-consistently made out of 
particles trapped around these loops, then it should be thinner while
perpendicular to the main bar.
Also, the $x_2$ loops lead the small bar as it moves from the
parallel to perpendicular position, and 
trail behind it when it moves back to align with the main bar.
One can conclude that in a self-consistent model, the small bar should
decelerate when going out of alignment with the main bar, and accelerate 
when on its way back to the aligned position. 
These conclusions will be confirmed by the loop structure of Model 2. 

In Model 1, the $x_2$ loop family extends only to $x$=2 when the bars 
align on the $x$-axis, and only the loops extending roughly to $x$=1.5 
there follow the small bar in its motion. But the secondary bar itself
extends to $a_S$=4.2 kpc, so there are no loops which
support the outer part of the secondary bar. By analogy with 
argument of Contopoulos \& Papayannopoulos (1980) for the $x_1$
orbits in a single bar, we conclude that a self-consistent 
secondary bar cannot extend outwards beyond the region where the
$x_2$ loops follow its motion relative to the large bar. The $x_2$ 
loops populate the region between the ILRs of the large bar. So we
conclude that the major axis of the small bar is unlikely to be 
larger than the outer extent of the 
$x_2$ orbital family of its main bar host along its major axis, unless
the strength of the secondary bar relative to the main one is large
enough to cause major changes in the loop structure.

\subsubsection{The $x_1$ loops}
\begin{figure}
\resizebox{11cm}{!}
{\includegraphics{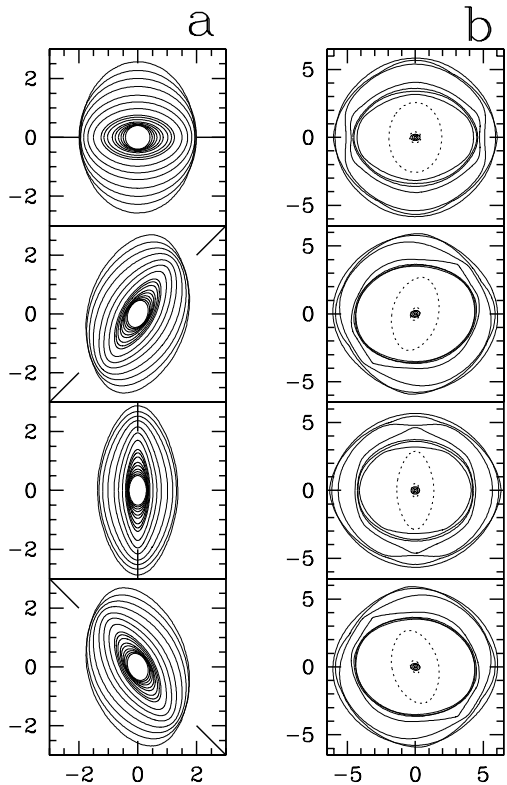}}
\vspace{-1cm}
\caption{{\it a)} The $x_2$ loops in Model 1 in the rotating frame of the 
large bar, which remains horizontal.  The straight lines at opposite 
sides or corners indicate the position angle of the small bar; time 
increases down the page. 
{\it b)} The $x_1$ loops at the same relative bar phases. The inner- 
and outermost $x_2$ loops are marked by dotted lines; three 
$x_{\rmn{1i}}$ loops are displayed inside them. Units are in kpc.}
\label{m1loops}
\end{figure}

The inner $x_1$ loops (the $x_{\rmn{1i}}$ subfamily) show a very regular 
behavior in Figure \ref{m1}a, where they are seen as a thick line in the lower
left corner. The $v_x(y)$ curve is continuous throughout the whole region
where these loops were detected: from the very center of the potential
to $y$=0.223. At this $y$, the $x$-velocity increases rapidly with $y$
and the loop family is lost in Figure \ref{m1}a. The $x_{\rmn{1i}}$ loops are 
almost circular and are represented by three innermost curves in Figure \ref{m1loops}b.

Unlike the $x_{\rmn{1i}}$ loops, loops belonging to the $x_{\rmn{1o}}$ 
subfamily (outer $x_1$ loops) do
not define a continuous function $v_x(y)$ in the plot of Figure \ref{m1}b.
The loop searching code repeatedly lost track of the minimum after
covering only a short $y$ range. For some $y$ ranges,
two comparably deep and wide minima were present, indicating that two
loops exist at the same $y$ but with different starting velocities $v_x$.
The appearance of the $x_{\rmn{1o}}$ loops is shown in Figure \ref{m1loops}b.
It is not regular, and these loops cover a very small part of the
primary bar. Therefore it appears unlikely that the primary bar can
be built in a self-consistent manner in Model 1.

\subsubsection{Other loops}
\begin{figure}
\resizebox{12cm}{!}
{\includegraphics{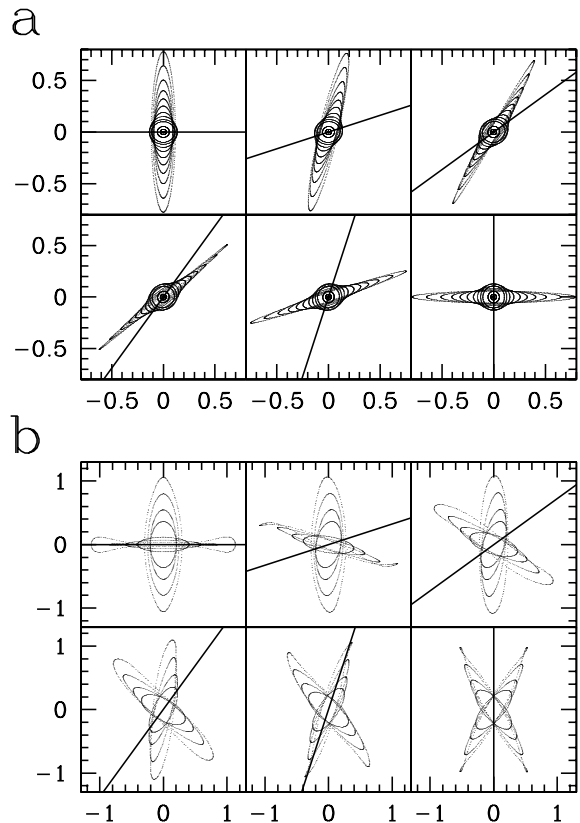}}
\vspace{-1cm}
\caption{{\it a)} Evolution of the representatives of the loop 
family marked by $A$ in Figure \ref{m1}, displayed in the 
rotating frame of the large bar, which remains horizontal. The 
straight line indicates the position angle of the small bar. Only
loops for the bars coming out of alignment are displayed: as
the bars come back to alignment, the loops are mirror reflections 
of these.
{\it b)} Representatives of the family marked by $B$ in Figure \ref{m1}
displayed in the same way as those in {\it a)}. Labels are in kpc.}
\label{m1otherloops}
\end{figure}

Although the secondary bar has an ILR, we found no loops corresponding 
to its $x_2$ orbits. Instead, we found a family of loops indicated by
the set of dark pixels in Figure \ref{m1}a (marked there by $A$), that departs to
the right from the thick band marking the $x_{\rmn{1i}}$ loops. Although
this family is not surrounded by a dark band of particles 
trapped around these loops, the minimum ring width is
well defined, and loops representative of this family
are displayed in Figure \ref{m1otherloops}a. When the two bars are 
aligned on the $x$-axis, the loops are perpendicular to them. But
the loops precess in a sense opposite to the small bar, aligning with it
when it makes an angle $\pi/4$ with the main bar, and becoming perpendicular
to the small bar again when it in turn is perpendicular to the main
bar. Thus the figure of the loop moves retrograde, but 
loop particles remain on prograde orbits with respect to either bar. 
The inner loops become rounder, and this family joins the $x_{\rmn{1i}}$ loops
at the point $y$=0.13, $v_x$=57 (Fig. \ref{m1}b). This family obviously 
does not correspond to the $x_2$ orbital family in the small bar,
and it does not support either bar. Nevertheless, it is consistently
present in further modifications of Model 1, and in Model 2 described
below.

Just as we can have multiply periodic orbits in a single bar, we can have
higher order loops in double bars. If particles come back to the same
curve only every other bar alignment, then after an odd number of 
alignments they lie along one
curve, and after an even number, on another curve. Thus a single particle
generates two closed curves in this case. One can easily extend this
to $N$ curves generated by a single particle returning to the same
curve after every $N$ alignments of the bars.
We found many such higher order loop families in our search, one
of which is presented in Figure \ref{m1otherloops}b:
both curves constituting each loop there precess retrograde, with a period
twice the the relative period of the two bars. In Figure \ref{m1}b, this
family (marked by $B$) runs parallel to the abovementioned family, when 
the larger
out of two $y$-axis crossing points is considered to be a loop generator.
Obviously, this family does not support any bar, and it
is very unlikely that any of the higher order loop families can do it.
Although we expect that abundant and complex higher order loops are
present, we did not search for them.
 
\section{Approach to a self-consistent doubly barred model}
Louis \& Gerhard (1988) have shown, and the previous section 
illustrates, that bars which make up a self-consistent double-bar system 
must distort slightly as they rotate through each other. Thus our
assumed potential, with two rigidly rotating bars, cannot be fully
self-consistent. Here we are looking only
for a model that may be close to being self-consistent, in which 
stable regular orbits give support to both bars in their motion.
A real bar would have some chaotic orbits as well, which we do not 
consider here.

\subsection{Choice of parameters leading to a nearly self-consistent model}
A system similar to Model 1 could not be built in a self-consistent manner: 
its $x_2$
loops do not extend to the end of the secondary bar, and the $x_1$ loops 
cover only a small fraction of the primary bar. Making the secondary 
bar stronger may help to extend the $x_2$ family, but it would destroy
the $x_1$ loops, which are already quite disrupted by the small bar
in Model 1. A different solution has to be found.

First, we modified Model 1 by increasing the $a/b$ axial ratio of 
both the bars from 2.5 to 4.0, while the product $ab$ of the axis lengths 
remained constant. The $x_1$ loop family virtually 
disappeared, except for the inner $x_{\rmn{1i}}$ loops. The 
$x_2$ family broke into two well separated parts, and it may be difficult 
to support the secondary bar as well. On the basis of our limited search, 
weakness of the $x_1$ family seems to be a general feature of models 
with a relatively large, fast-rotating inner bar, like Model 1.
 
The new model, hereafter Model 2, is meant to be as close to 
Athanassoula's (1992a) standard model as possible. The quadruple moment 
of the primary bar is twice as large as in Model 1. From Model 1, 
we know that the secondary bar should not be too big. First we 
tried a long, but not massive ($a_S = 0.4 a_B$, $M_S/M_B = 0.2$),
secondary bar. The phase plane search showed that the
$x_1$ family, which should support the longer bar, is practically 
absent for $y>$0.9, while the bar length is $a_B = 6$. The $x_2$
family is broken into two parts, both surrounded by regions of
trapped semi-periodic orbits. This model may have a 
nearly self-consistent secondary bar, but cannot support the primary bar, 
and therefore has to be rejected.
 
Then we tried a smaller, but more massive secondary bar 
($a_S = 0.2 a_B$, $M_S/M_B = 0.3$). The loops from the 
$x_1$ family extend far enough in radius ($1.1<y<4$)
to support part of the primary bar. The inner part of the 
$x_1$ family is quite extended, though: it suppresses the $x_2$ family,
which in turn appears irregular and disrupted on the phase plane. In 
this model, part of the primary bar can be supported by loops,
but the secondary bar is likely to be far from self-consistent.

The two attempts above suggest that our secondary bar is still too massive, 
or too large. Observations presented in \S4 confirm this suspicion:
the ratio of semimajor axes of the bars in the model, is beyond the 
observed range for our model with $a_S=0.4a_B$, and at the lower end 
of the observed range for the model with $a_S=0.2a_B$.
Guided by the values listed in Table 2, we decided to make 
the secondary bar smaller and less massive. The ellipticity of the 
observed secondary bars is considerably smaller than that of the primary
ones;
therefore we set the axial ratio of the secondary bar at $b_S/a_S=0.5$.
The small bar is only 20\% as long as the primary bar and
has only 15\% of the big bar's mass. Since in Model 1 the loops supporting
the small bar failed to extend all the way to its corotation, in Model 2
we set the pattern speed so that
the Lagrangian radius of the secondary bar is at $1.9a_S$. This also 
puts corotation of the small bar at the outer ILR of the big one, as 
suggested by Tagger \etal (1987).
 
\begin{figure}
\vspace{-2cm}
\resizebox{7cm}{!}
{\includegraphics{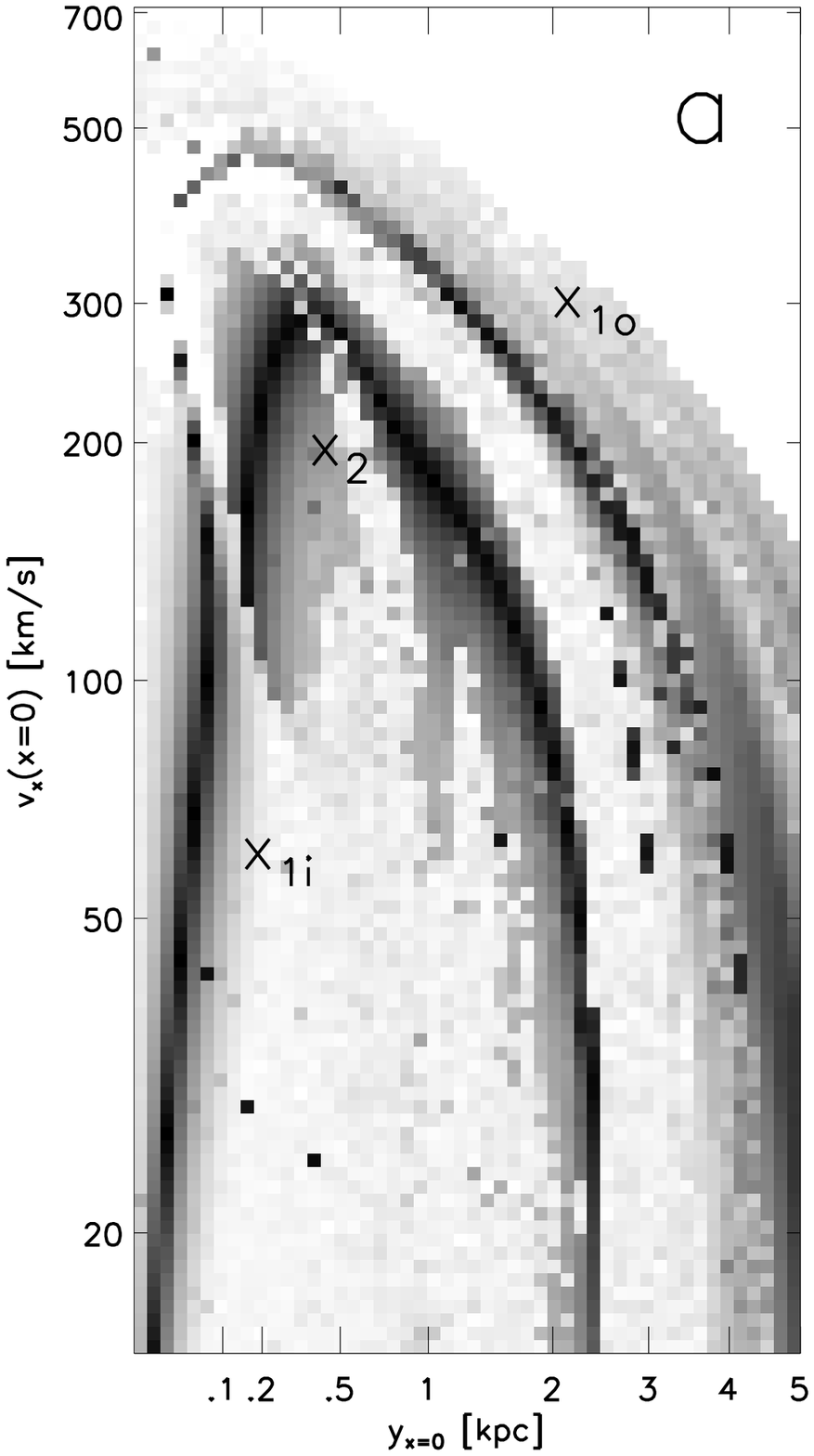}}
\resizebox{9cm}{!}
{\includegraphics{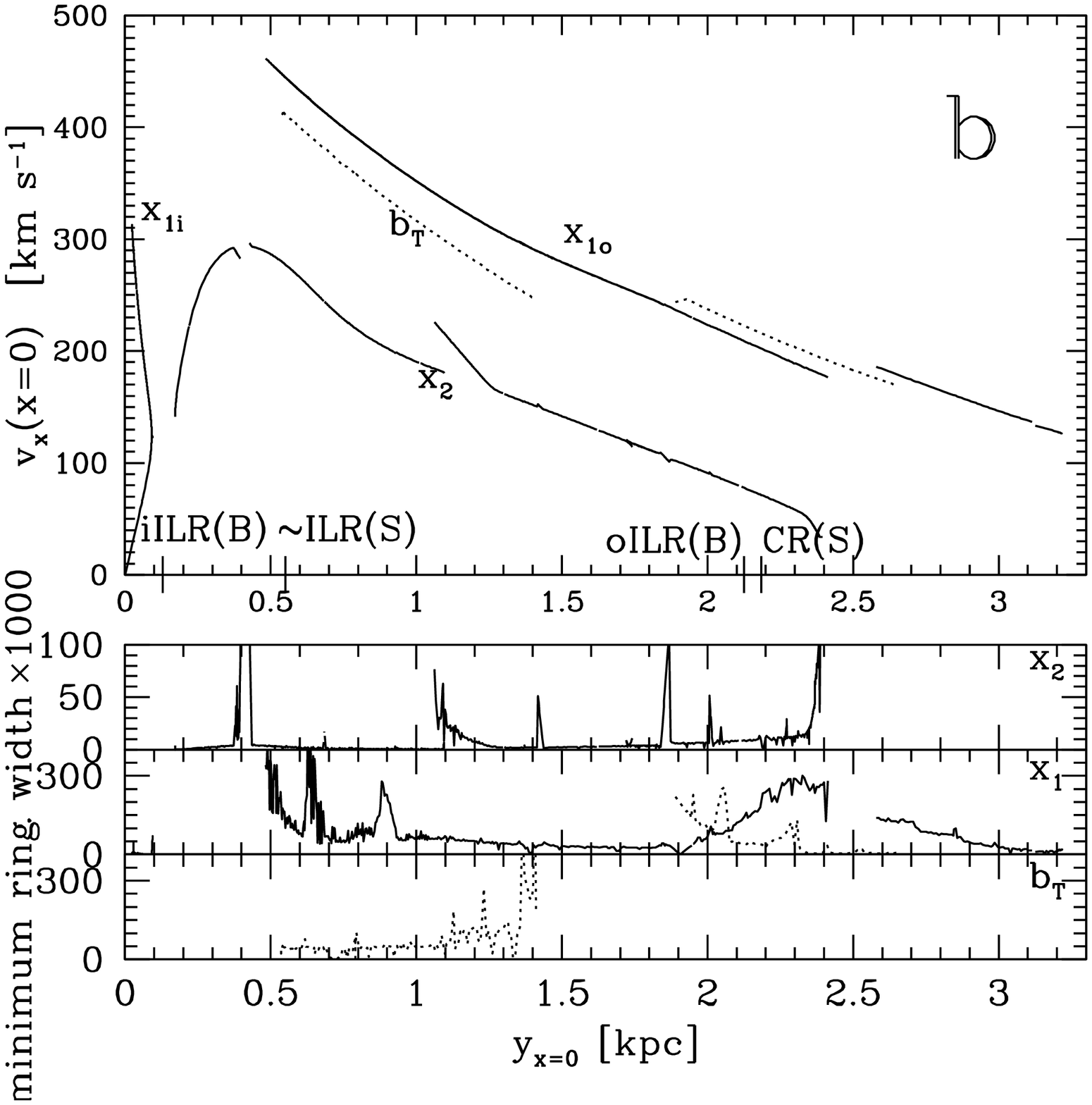}}
\vspace{-.1cm}
\caption{{\it a)} Phase plane search for Model 2 with two bars initially
on the $x$-axis. {\it b)} The characteristic curves for Model 2 
at bars aligned. The diagrams of minimal ring widths $w(y)$ along each 
family are attached. For clarity, some characteristic curves and the
corresponding widths are marked in dotted. Note that the widths change 
much more abruptly with $y$ than in Figure \ref{m2sgl}.}
\label{m2}
\end{figure}

\subsection{Loop families supporting a nearly self-consistent Model 2}
We constructed Model 2 with a particular interest in loops supporting
the potential. We want to answer the question, is it possible
to construct a model close to this one in a self-consistent way? 
Our first task is to survey the structure of $x_1$ and $x_2$ loop 
families,
which we expect to support the primary and secondary bar, respectively.
The results of the phase plane search for Model 2
are displayed in Figure \ref{m2}a. Both $x_1$ and $x_2$ loop families are
present there, with trapped orbits around them indicated by the
dark areas. Note how similar this figure is to Figure \ref{m2sgl}a, in which
a phase plane search for Model 2 with the main bar only was presented.
Unlike in Model 1, this plane search suggests that both 
bars are likely to be supported by stable loops. Figure \ref{m2}b shows
the characteristic curves and the variation of minimum ring width
$w(y)$ along each loop family. We find an additional loop family, marked
there by $b_T$, supporting part of the main bar.

\subsubsection{The $x_2$ family --- loops supporting the small bar}
We were able to find loops belonging to the $x_2$ family 
for $0.17<y<2.39$. Just before the $x_2$ family is lost at 
small $y$, the velocity drops rapidly. At the high-$y$
end of the $x_2$ family, the ring widths increase making it harder to 
find the loop, before the loops 
disappear (Fig.\ref{m2}b). This family has two interesting features.
First, around $y$=0.406, the characteristic curve is discontinuous, and
there is a small $y$ range where no loops have been found. This 
corresponds to the white stripe crossing the $x_2$ arch in the phase 
plane search of Fig. \ref{m2}a. The minimal ring width $w(y)$ increases steadily 
as this discontinuity is approached from both lower and higher $y$'s,
before the width experiences a sudden jump. A detailed search for loops 
in this $y$ range found no simple smooth invariant 1-D curve; but for a 
wide range of $v_x$, the particles are confined to well-defined rings. 
No higher order loops were detected in this region. It is possible that a
similar small discontinuity occurs at $y$=1.86. Here, the ring width 
jumps an order of magnitude above its normal level over a narrow
$y$-range, where we find no simple smooth 1-D curves. Again, for a wide 
$v_x$ range we observe that the particles are confined to rings there.

\begin{figure*}
\vspace{-2.5cm}
\resizebox{23cm}{!}
{\includegraphics{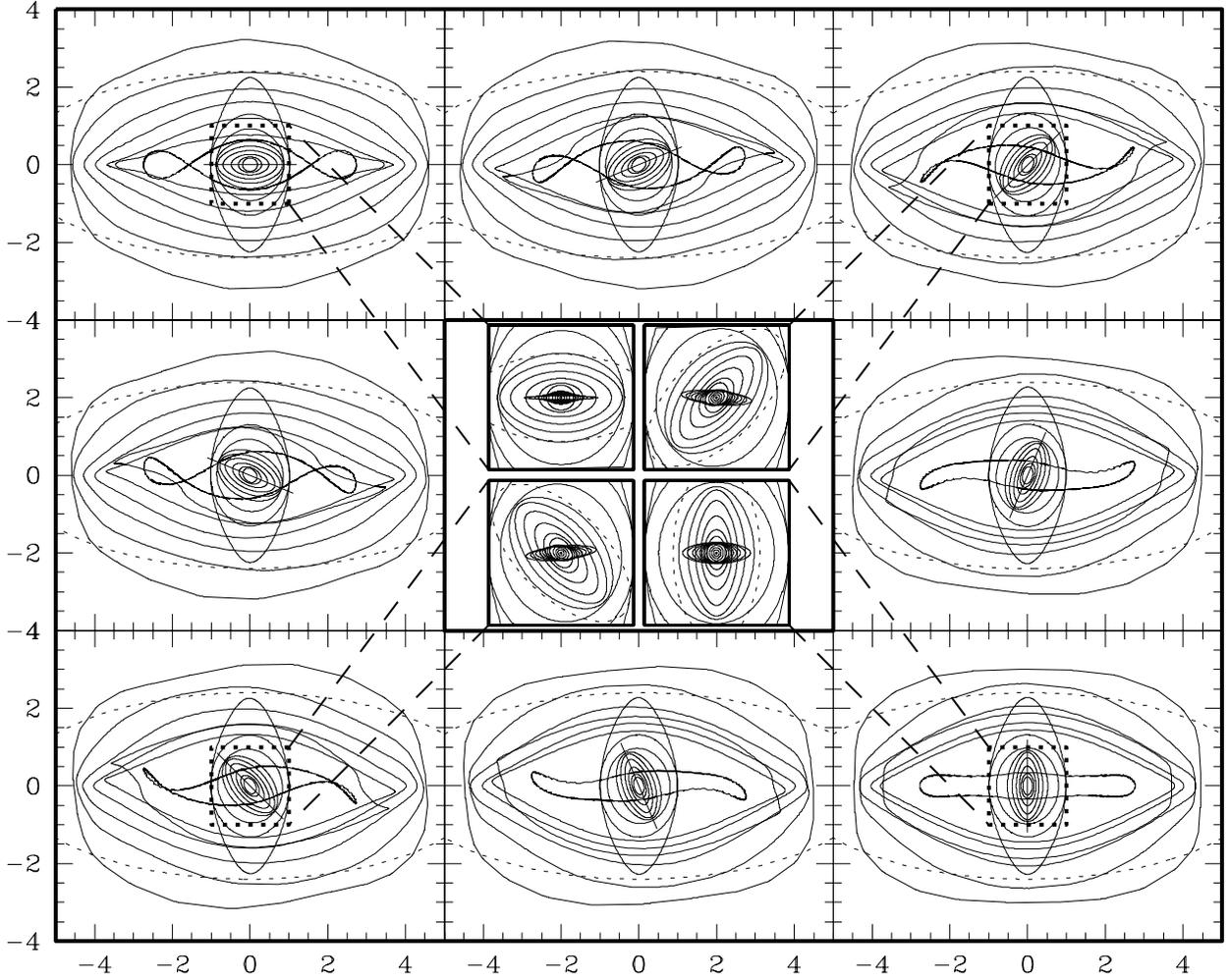}}
\vspace{-7cm}
\caption{An overview of loops in a nearly self-consistent Model 2 of 
a doubly barred galaxy. The loops are displayed in the reference
frame of the main bar at different relative positions of the bars. 
The outer bar is outlined by the dashed line and the major axis of 
the secondary 
bar is marked by a straight line on outer panels. The sequence follows 
along outer panels clockwise, with central regions magnified on inner 
panels, where the secondary bar is outlined by the dashed line. 
Units on the axes are in kiloparsec.}
\label{m2loops}
\end{figure*}

At $y\approx1.08$, the $v_x(y)$ curve for the $x_2$ loops splits 
in two; over a small range,
there are two loops for each $y$ value. The accompanying plot of ring
widths shows that when approaching this region from higher $y$, the 
width of loops increases rapidly, and the searching code effectively 
loses the loop family. The minimum followed by our continuous search is no
longer the global minimum, and another minimum is found which generates
the continuation of this family to smaller $y$.
 
The overall appearance of the $x_2$ loops in Model 2 is presented in 
the central panels of Figure \ref{m2loops}. The $x_2$ loops with $y\la0.5$ rotate 
with the small bar and extend out to about 0.85-0.9 of its semimajor 
axis. There the density in an $n$=2 Ferrers bar has dropped to 4\% of 
its central value. Thus we
conclude that the secondary bar is well supported by $x_2$ loops: even
if the family breaks at $y=0.4$, there are `near-loop' solutions, and 
particles trapped around these can serve as the building blocks of 
the inner bar. Here we can also confirm our conclusion drawn on the 
basis of Model 1, that the secondary bar pulsates and accelerates,
although these effects are smaller in Model 2.

\subsubsection{Loops supporting the larger bar --- the $x_1$ and $b_T$ 
families}
Figure \ref{m2}b shows two disconnected parts of the $x_1$ family, which were 
detected with the search method described in section 3.2: the 
$x_{\rmn{1i}}$ loops (inner
$x_1$) at $y<0.09$ and the $x_{\rmn{1o}}$ loops (outer $x_1$) at $y>0.5$.

The $x_{\rmn{1i}}$ loops show a rapid growth in the $x$-velocity
with $y$, and for $v_x>$120, this loop family continues for decreasing 
$y$ with velocities still growing. Therefore the $x_{\rmn{1i}}$
loops are double-valued in this range. They cannot be traced beyond the
point $v_x$=310, $y$=0.024, corresponding to a dark pixel in Figure \ref{m2}a.
We explore the question of connection between the inner and outer $x_1$ 
families later in this section.

The $x_{\rmn{1o}}$ loops are not as well defined as the $x_{\rmn{1i}}$ 
and $x_2$ loops --- the minimum ring widths are larger. Near
$y$=0.65 and 0.85, they rise to unacceptable values of 50\% and 30\% of 
the ring size, but no discontinuity in the loop family tracing is seen.
We will return to this feature later in this section. In general, for the
$x_{\rmn{1o}}$ loops there is no clear single $w(v_x)$ minimum for a 
given $y$, but rather several less deep minima, of which one usually 
dominates. This is similar to what the $x_2$ loops experience near
$y=1.08$. For $x_{\rmn{1o}}$ loops, two minima
of roughly equal depth persist for the $y$ range from 1.85 to 2.4,
and are displayed as two parallel lines 
in Figure \ref{m2}b. In this region, the ring widths increase with $y$ for 
the loops marked with a solid line, and this branch of the $x_{\rmn{1o}}$ 
loops is eventually lost at $y$=2.4. At that $y$, the widths on 
the other branch, marked with a dashed line, become small and behave in a 
very regular way. This branch is lost in turn at $y$=2.65. 
At 2.57$<y<$3.22, another branch of the $x_{\rmn{1o}}$ loops exists;
this is the branch extending to the highest $y$ in Model 2. We consider
the solid line extending from $y$=0.5 to 2.4 to be the main part of
the $x_{\rmn{1o}}$ family, and we call the two abovementioned branches
the $x_{\rmn{1o}}$ family additions.
 
The dotted line parallel to the $x_{\rmn{1o}}$ family and persisting 
for the $y$ 
range from 0.54 to 1.4 (Fig. \ref{m2}b), marks another loop family found in 
Model 2, which we name $b_T$ for their 'bow-tie' appearance. Below 1 kpc, 
the ring widths for the $b_T$ loops are smaller than those of the 
$x_{\rmn{1o}}$ loops, and the widths behave in a more regular way. 
 
Figure \ref{m2loops} displays representatives of all loop families found for
Model 2, as they change while the two bars rotate relative to
each other. The two outermost loops belong to the two 
outer additions to the $x_{\rmn{1o}}$ family. The next four loops inside 
them belong to the main part of the $x_{\rmn{1o}}$ family, and they 
undergo interesting oscillations when the two bars rotate through one 
another. The loop that is innermost when the bars are aligned 
(top left panel) develops cusps, and becomes much rounder: it ends up 
as a second outermost loop when the bars are orthogonal
(bottom right panel). This leaves us with no $x_{\rmn{1o}}$ loops 
supporting the inner part of the main bar when the bars are orthogonal 
(bottom right panel). But there is one more loop inside the above mentioned
set of four, which remains parallel to the main bar. It intersects itself 
at bars aligned, and belongs to the $b_T$ family marked in Figure \ref{m2}b by a 
dotted line. Loops from this family take a boxy shape at bars orthogonal
and fill the central part of the main bar, thus supporting it in addition
to the $x_1$ loops.
 
The $x_{\rmn{1i}}$ loops are displayed against the $x_2$ loops in the 4 
inner panels of Figure \ref{m2loops}. The double-valued $v_x(y)$ curve for
these loops means that when the two bars are aligned,
two loops from this family cross the $y$-axis at the same point. One 
set of these loops is close
to circular and can be detected all the way to the potential center, the
other one is elongated and gets longer with decreasing $y$, up to a
semimajor axis of 0.5 kpc for $y$=0.024, where the family disappears.
These narrow loops get stretched when the bars are orthogonal, but their 
semimajor axes roughly do not change, and they remain well aligned with 
the primary bar. Note that the $x_{\rmn{1i}}$ loops are narrowest at
bars aligned, unlike the $x_2$ loops which are narrower at bars
orthogonal. Figure \ref{m2loops} also shows that 
in a potential of two independently rotating bars, sets of loops 
supporting  the bars can penetrate through one another while each 
follows a different bar. If there are enough stars trapped on the
$x_{\rmn{1i}}$ loops, such a galaxy may appear as having a triple bar, with 
innermost and outermost bars aligned.

\begin{figure}
\vspace{-2cm}
\resizebox{7cm}{!}
{\includegraphics{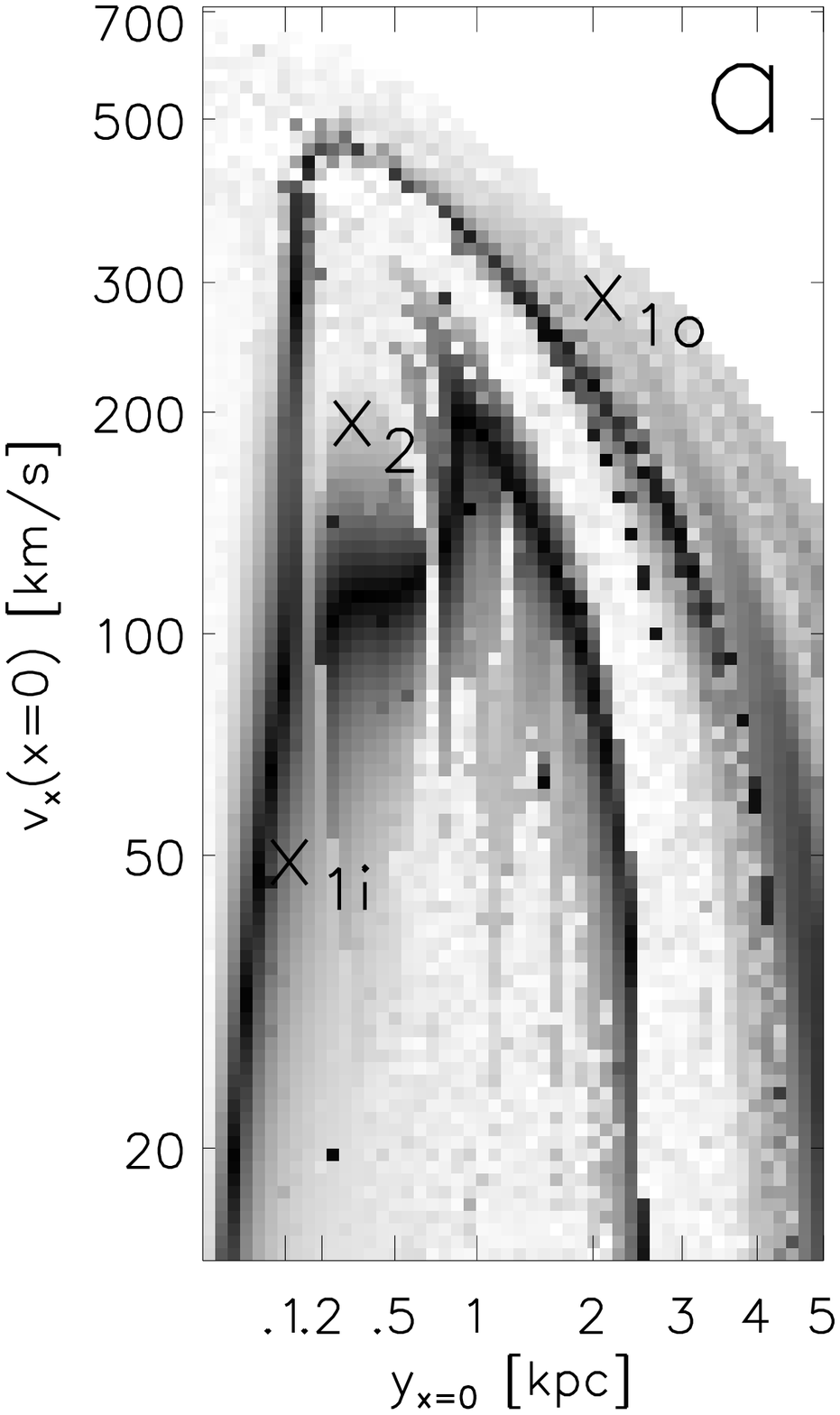}}
\vspace{-4cm}

\resizebox{7cm}{!}
{\includegraphics{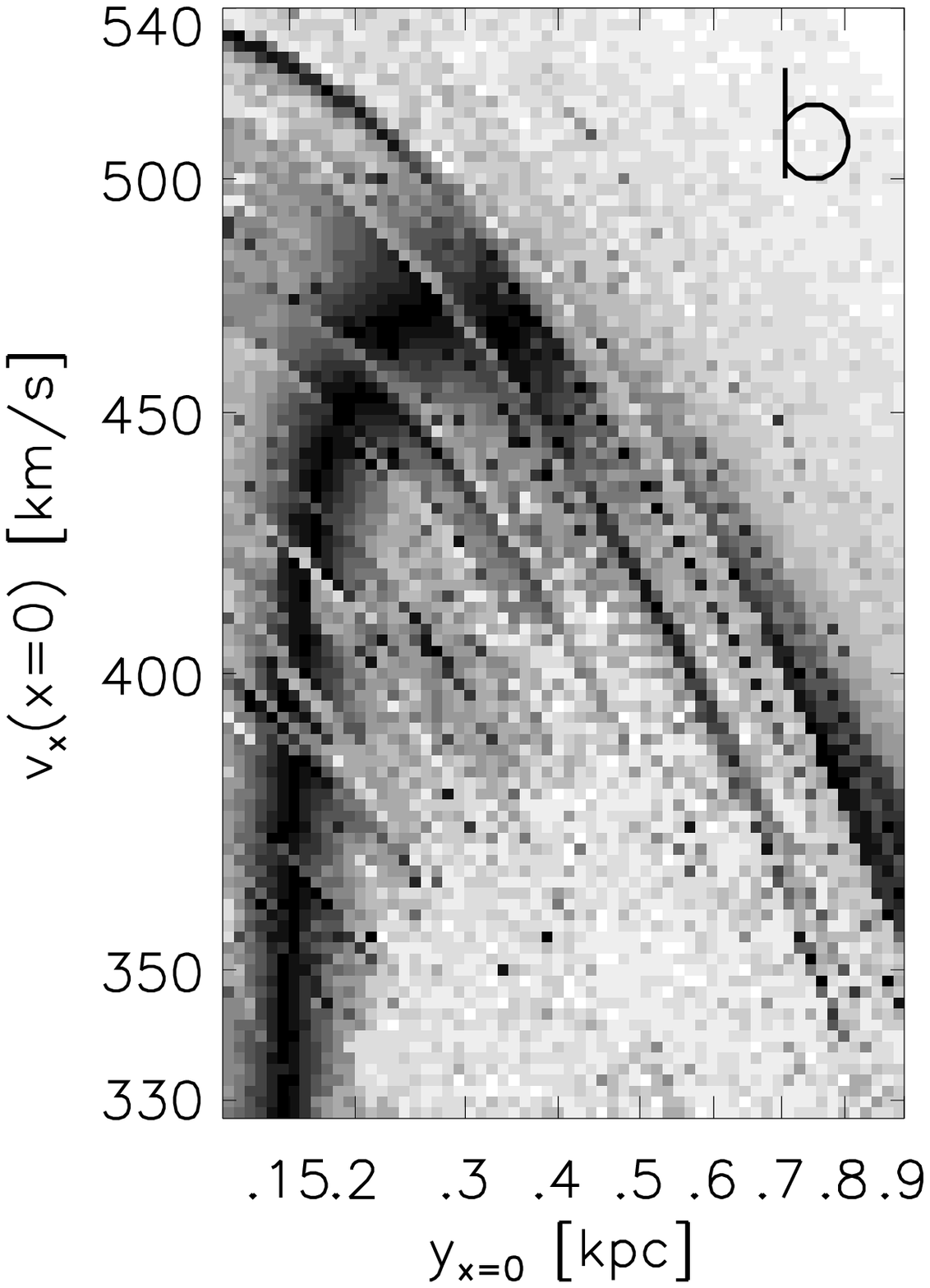}}
\vspace{1cm}
\caption{{\it a)}As Fig.\ref{m2} but for a particle starting
on the minor axis of the main bar when the bars are orthogonal. {\it b)}
Close-up of the top of the `$x_1$ arch' from {\it a)}.}
\label{m2ort}
\end{figure}
 
\subsubsection{Loop search with bars orthogonal}
One can get a more complete insight into the loop families in double bars
by performing an additional search, with particles starting on the
minor axis of the main bar (the $y$-axis), moving perpendicular to 
that axis, when the bars were orthogonal. Phase plane search of Figure 
\ref{m2ort}a indicates both the $x_1$ loops (the upper 
arch) and $x_2$ loops (inner arch and plateau). A close-up of the upper 
arch is shown in Figure \ref{m2ort}b: many bright strips running from 
the top-left 
to the bottom-right indicate areas where the iterates are no longer 
confined to a ring. The $x_1$ family is broken when they cross its
dark arch. The structure at such intersections is similar to the
features common in characteristic diagrams for orbits in a fixed 
potential, usually called the 4/1-like gaps (\eg Contopoulos \&
Grosb{\o}l 1989). It can be best seen at $y=0.2$, $v_x=460$ in Figure
\ref{m2ort}b. Around $y=0.6$, there is a white strip which separates two dark 
strips: one surrounding the $b_T$ family and originating at the
top-left corner, and the other surrounding the $x_{\rmn{1o}}$ loops
originating on the right. It looks like this white strip overlaps
with what would be an $x_1$ orbital family in a single bar there
(see Fig. \ref{m2sgl}a), and it is unlikely that stable loops can be found there.

\begin{figure}
\resizebox{9cm}{!}
{\includegraphics{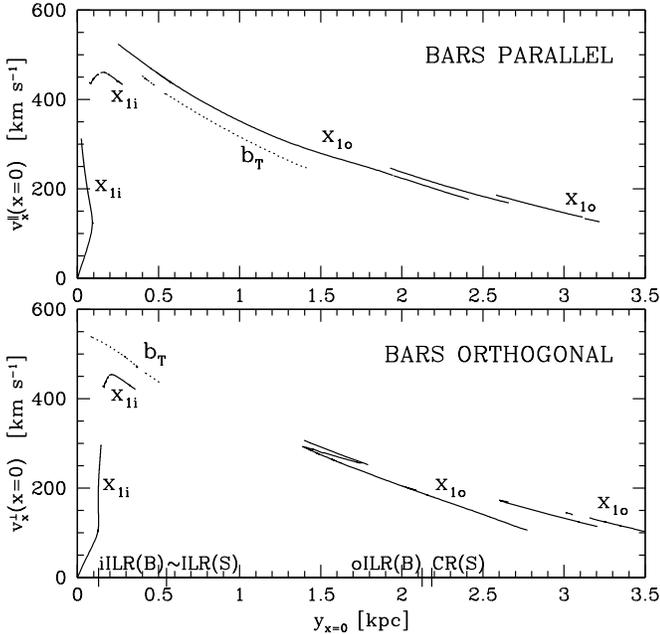}}
\vspace{-.1cm}
\caption{Two diagrams to compare the characteristic curves of loop 
families supporting the main bar in Model 2 at bars aligned ({\it top}) 
and orthogonal ({\it bottom}).}
\label{m2comp}
\end{figure}

Comparing the loop families found with particles starting at the 
bars orthogonal with those found by the original search method earlier 
in this section allowed for
a more complete picture of loops supporting the main bar. Figure \ref{m2comp}
shows the characteristic diagrams for these loops, both at bars aligned
and orthogonal. The $x_{\rmn{1o}}$ loops, which occupied the region from 
0.22 to 2.4 kpc in $y$ at bars aligned, are compressed into a $1.4<y<2.8$ 
range at bars orthogonal. In this configuration, we find no 
$x_{\rmn{1o}}$ loops at $y<1.4$ to support the main bar. When the bars
are orthogonal, at 1.4$<y<$2.2 this family 
develops an interesting zig-zag on the characteristic curve 
(Fig.\ref{m2comp}b). It means that the innermost loops at bars aligned 
get pushed away at bars orthogonal and end up as the outermost ones. We 
gave an example of such a loop in section 6.2.2.
On the diagonal of the `Z' the trend reverses, and loops pulsate without 
changing their order. Eventually, the trend reverses again (the top of 
the `Z'); the loop which is outermost when the bars are aligned sinks 
inside, and is innermost at bars orthogonal. At the points ($y=$1.4, 
$v_x^{\perp}$=308) and ($y$=1.8, $v_x^{\perp}=$250), where the 
$x_{\rmn{1o}}$ characteristic curve reverses, there is a set of
rings that for a short range continues in the direction before
reversing, but the ring widths increase rapidly.
 
In some regions of the lower 
branch of the `Z' at bars orthogonal, loops do not transform to points on 
the $x_{\rmn{1o}}$ characteristic curve at bars aligned monotonically,
but get spread over a large $y$ range. 
The transformed curve is disrupted at several places, which correspond to 
high ring widths for loops found at bars aligned (Fig.\ref{m2}b). Some real 
gaps in the stable loop families likely
occur there, which are not just artifacts of our method. Thus the
`high width' found with our method may suggest that the loop sequence is
broken and we may encounter part of a different family with a short 
stability range.

The $b_T$ family and the two disconnected $x_{\rmn{1o}}$ branches at high 
$y$ in Figure \ref{m2comp}a
reverse order at bars orthogonal (the innermost loop becomes the outermost
one). The two outer additions to the $x_1$ family may be connected by bridges
similar to the diagonal of the `Z' in the main $x_{\rmn{1o}}$ subfamily:
at $y=2.6$, $v_x^{\perp}=170$, the first addition to the $x_{\rmn{1o}}$
loops changes order and aims towards the second addition, but we were not
able to find a continuation for this set of loops.

We tried to find loops that could connect the $x_{\rmn{1i}}$ and $x_{\rmn{1o}}$
subfamilies. Our search at bars orthogonal resulted in finding loops 
represented by an arch at $y$=0.2, $v_x^{\perp}=$450 in Figure \ref{m2comp}b.
It may be a continuation of the $x_{\rmn{1i}}$ subfamily extending to the 
galaxy center, but no connection of those two has been found. Several
gaps in the `$x_1$ arch' there seen in the phase plane search (Fig.\ref{m2ort}b),
suggest discontinuities in the $x_1$ family.
The $x_{\rmn{1i}}$ arch in Figure \ref{m2comp}a most likely 
has no connection with $x_{\rmn{1o}}$ loops either, since the $b_T$ loop
family seems to separate them.

\subsubsection{Summary of loops supporting the main bar in Model 2}
The main $x_1$ family in Model 2 consists of 2 subfamilies with some additions:
\begin{enumerate}
\item the inner $x_{\rmn{1i}}$ subfamily originating at ($y=0$, $v_x=0$),
\item the addition to the inner $x_{\rmn{1i}}$ loops marked by the arch
in Figure \ref{m2comp},
\item the main $x_{\rmn{1o}}$ subfamily: there is no continuous transition 
from this subfamily to the $x_{\rmn{1i}}$ subfamily,
\item two outer additions to the $x_{\rmn{1o}}$ subfamily.
\end{enumerate}
There is an additional family, the $b_T$ or bow-tie loops, but it 
still leaves a gap 
without a continuous set of simple stable loops for 0.5$<y<$1.4 at 
bars orthogonal. There are some small loop subfamilies found at 
0.7$<y<$1.5 at bars orthogonal, but they are discontinuous. There 
are absolutely no loops at 0.5$<y<$0.7 --- there is a white strip of 
diverging iterates overlapping that part of the dark `$x_1$ arch' on 
the phase space diagram in Fig.\ref{m2ort}b.
 
In conclusion, we would say that the $x_1$ and $b_T$ loops support most of
the main bar at bars aligned. The $x_{\rmn{1o}}$ loops are strongly 
influenced by the motion of the secondary bar. At bars orthogonal, they
get rounder and group in the outer parts of the main bar. We were not 
able to find any $x_1$ loops supporting the inner part of the main bar 
when the secondary bar is perpendicular to it --  the $b_T$ loops fill 
up the center there, but a small gap between these two families remains 
(Fig.\ref{m2loops}). In section 6.2.1 we showed that the secondary bar is fully
supported by the $x_2$ loops.

\section{Discussion, consequences for gas flows}
The main goal of this work was to explore whether 
a galactic stellar disk with two non-equal bars rotating 
at different pattern speeds is dynamically possible. The `backbone'
for the orbital structure of this time-periodic system
is provided by stable loops. These are closed curves, such that 
particles placed on them and moving in the potential of this 
system will return to the original curve after two bars have 
come back to the same relative orientation. A nearly self-consistent
double bar should have stable loops elongated with each bar over
appropriate radial range. In section 6, we gave an example of a doubly 
barred
system where stable loops existed with the right shape, over the right 
range in radius, to support both the inner and the outer bars.

To create conditions most favorable to the existence of stable loops,
we selected a smooth potential by choosing a high Ferrers index of 
the bars ($n=2$). We also minimized the number of resonances by putting 
the corotation of the small bar at the outer ILR of the main bar. 
This overlapping of resonances, although not essential for the
existence of self-consistent double bars, may be natural in dynamical
systems (see Tagger \etal 1987). Our secondary bar was slow rotating,
with corotation well beyond end of bar. To have the same resonant
condition for a fast-rotating secondary bar, the inner bar would have 
to extend to the outer ILR of the main bar. In such a case, all the
$x_2$ loops should follow the secondary bar in its motion, a
configuration that we were not able to create. On the basis of 
our models, we believe that for double bars 
we may not be able to have both resonance overlapping and a secondary 
bar extending to its corotation.

The morphology of the gas flow in a barred potential strongly depends 
on the bar pattern speed. Straight shocks, often seen in large-scale
bars, occur when the bar is rapidly rotating, \ie it extends at least 
to 80\% of its corotation. For slower bars, the shocks curl around the 
bar and start forming a ring (Athanassoula 1992b). Thus in a slowly
rotating secondary bar, we expect an elliptical flow to develop 
instead of straight shocks (Maciejewski 1999). The actual gas morphology 
will obviously
be influenced by stars-gas interaction, self-gravity in the gas, and
star formation processes, none of which is considered in this
simple scenario.

\section{Summary}
We have shown that doubly barred galaxy potentials allow stable 
regular orbits that can support the shape of both inner and outer
bars. These multiply barred galaxies can be built as gravitationally
self-consistent systems. In doubly barred galaxies, the most 
important loop families, which serve as their backbones, occupy the 
same parts of phase space as the $x_1$ and $x_2$ orbits in the single
bar. The loop family corresponding to the $x_1$ orbital family in the 
main bar, always remains elongated along the main bar, even though 
the inner loops reside inside the secondary bar.
The $x_2$ loop family, which at large radii behaves like  
the $x_2$ orbital family of the main bar, follows the 
secondary bar in its motion at smaller radii.
No loop family corresponding to the $x_2$ orbital family in the
secondary bar has been found. 

Despite a limited number of models, we were able to draw conclusions 
about constraints that the orbital structure puts on the shapes, 
relative sizes and frequencies of the bars.  
The secondary bar can be supported by the $x_2$ loops which rotate 
smoothly with its figure. These loops change axial 
ratio as the bars rotate, and lead or trail the figure of the
secondary bar, depending on the relative phase of the two bars.
A self-consistent secondary bar, supported by stars trapped around
these loops, must pulsate and accelerate as it revolves inside the
main bar. Since the $x_2$ loops originate from the $x_2$ orbits in
the potential of a larger bar, the size of 
a self-consistent secondary bar is approximately
limited by the maximum extent of the $x_2$ orbits along the main bar's 
major axis. Because of these limitations, we were not able to find a 
self-consistent model of a doubly barred galaxy in which the secondary 
bar remains in resonant coupling with the main bar, and extends all the 
way to its corotation. Our nearly self-consistent model contains a slowly 
rotating secondary bar.

A strong secondary bar can easily disrupt 
the $x_1$ loops supporting the main bar; thus there are upper 
limits for the mass and size of the small bar.
The inner region of the large bar can also be supported by
another loop family, which we call the bow-tie or $b_T$ family.
The loops supporting one bar rotate through the ones supporting
the other. Stars can be trapped around both these sets of loops, but 
gas can reside on at most one of them. 

The concept of the loop developed in this paper provides a powerful 
tool to study particle orbits in any periodically pulsating 
gravitational potential: in double bars, pulsating spheres, and eccentric
binaries. It is most promising in the last case: the motion of a massless 
test particle under the gravitational force of two point masses (the 
restricted three body problem), where only a few isolated closed orbits
have been found so far.

\section*{Acknowledgments}

We thank Peter Erwin for useful discussions and help with fitting 
isophotes to our models. This research was supported by NSF Extragalactic 
Program grant AST93-20403 and by NASA Astrophysics Theory Program grant 
NAGW-2796.

\onecolumn

\appendix

\section{Loops in a weakly barred potential: the epicyclic solution}
Here we show how approximations to loops can be found analytically within 
the epicyclic formalism. We also present the solution of a damped
epicyclic approximation, which yields excellent approximation to the
streamlines of gas flow in a singly-barred potential (Lindblad \& Lindblad
1994, Wada 1994). The predictions for gas flow in doubly 
barred galaxies were explored by Maciejewski \& Sparke (1997).

In the epicyclic approximation the motion of a test particle
can be decomposed into the `guiding center' motion, on
a circle of radius $R_0$ with the  angular velocity $\Omega(R_0)$
corresponding to circular motion in the axisymmetric potential $\Phi_0$, 
together with the 
epicyclic oscillations resulting from the forcing term $\Phi_1$, and a 
free oscillation at the local epicyclic frequency $\kappa (R_0)$. 
In a single bar, this formalism gives the approximate
orbits that are the counterparts of the $x_1$ and $x_2$ families.
On these closed orbits, the free oscillation is absent. 
In the same way, the solution of the epicyclic approximation in a double bar
can lead to finding counterparts of the main loop families.
We assume the free oscillation to be absent in the double bar as well.
We extend the epicyclic solution for orbits in a barred potential (see \eg
Sellwood \& Wilkinson 1993, Wada 1994 and Lindblad \& Lindblad 1994) 
to double bars.

\subsection{The epicyclic solution for a Hamiltonian System}
In planar polar coordinates ($R,\varphi$) the linearized equation of motion
(\ref{eqmot}) yields following equations for $R_1$ and $\varphi_1$ first 
order corrections to the guiding center motion
\begin{eqnarray}
\label{eqR1}
\ddot{R_1} + \left(\frac{\partial^2\Phi_0}{\partial R^2}_{|R_0} 
- \Omega_0^2\right)R_1 - 2 R_0 \Omega_0 \dot{\varphi_1}  & 
= & -\frac{\partial \Phi_1}{\partial R}_{|R_0,\varphi_0} , \\
\label{eqPhi1}
R_0^2 \ddot{\varphi_1} + 2 \dot{R_1} R_0 \Omega_0  & 
= & -\frac{\partial \Phi_1}{\partial \varphi}_{|R_0,\varphi_0} ,
\end{eqnarray}
where $\Omega_0=\Omega(R_0)$.
If in addition to the main bar rotating rigidly with angular velocity 
$\Omega_B$, there is another, secondary bar rotating
with angular velocity $\Omega_S$, the first order term $\Phi_1$ 
in the multipole expansion of the 
potential in the frame of the main bar becomes 
time-dependent, and consists of two bisymmetric terms
\begin{equation}
\Phi_1(R,\varphi,t) = \Psi_B(R)\cos 2\varphi + \Psi_S(R)\cos(2\varphi-\omega_p t) ,
\end{equation}
where $\omega_p = 2 (\Omega_S - \Omega_B)$, and the time $t$ was chosen such 
that $t=0$ when bars are parallel. The first term on the right-hand side 
corresponds to the main bar, at rest in its own frame, and the second one
describes the secondary bar, which rotates in this frame. After substituting 
$C_{B,S}=-(d\Psi_{B,S}/dr)_{|R_0}$ and $D_{B,S}=-(2\Psi_{B,S}/R)_{|R_0}$,
the right-hand sides of equations (\ref{eqR1}) and (\ref{eqPhi1}), for 
radial and azimuthal motion respectively, become
\begin{eqnarray}
\label{dblezeta}
-\frac{\partial \Phi_1}{\partial R}_{|R_0,\varphi_0} & = &
C_B \cos 2 \varphi_0 + C_S \cos(2\varphi_0-\omega_p t) , \\
\label{dbleeta}
-\frac{1}{R_0}\frac{\partial\Phi_1}{\partial\varphi}_{|R_0,\varphi_0} &= &
-D_B \sin 2 \varphi_0 - D_S \sin(2\varphi_0-\omega_p t) .
\end{eqnarray}
In general, we can start a particle orbit at any relative position
of the bars, at an arbitrary time $t_s \neq 0$, with its guiding
center at an arbitrary angle $\varphi_{0s} \equiv \varphi_0(t=t_s)$.
The angle 
$\varphi_0$ of the epicycle center at a given time $t$ is then
$\varphi_0 = \varphi_{0s} + (t-t_s)\frac{\omega}{2}$, where 
$\omega \equiv 2 ( \Omega_0 - \Omega_B)$. The two bars induce
two characteristic forcing frequencies, so
we introduce two time counters, to integrate the equations of
motion in a way analogous to Wada (1994). The first 
frequency is $\omega$ and the corresponding time $t'$ is defined by 
$\omega t'=2\varphi_0$. The second 
frequency is $\Delta \omega = \omega - \omega_p$ with the corresponding 
time defined by $\Delta \omega t'' = 2\varphi_0 - \omega_p t$. Equation 
(\ref{eqPhi1}) for the azimuthal motion, with right-hand side stated
explicitly in (\ref{dbleeta}), can be integrated over time to single out 
$\dot{\varphi}_1$ and substitute it back to equation (\ref{eqR1})
for the radial motion. Thus we 
have the second order equation for $R_1$ in the doubly barred case 
\begin{equation}
\label{dblezetabis}
\ddot{R_1} + \kappa^2 R_1 =
\frac{m_B}{\omega}\cos\omega t' +
\frac{m_S}{\Delta\omega}\cos\Delta\omega t'' ,
\end{equation}
where $m_B=\omega C_B + 2 D_B \Omega_0$,
$m_S=\Delta\omega C_S + 2 D_S \Omega_0$,
$\kappa$ is the free epicyclic frequency defined by 
$\kappa^2 \equiv 4\Omega_0(\Omega_0-A)$, and $A$ is the Oort constant.
In equation (\ref{dblezetabis}),
one can substitute $R_1 = R_{1B} + R_{1S}$ and perform
a standard separation of variables, which leads to two decoupled equations
\begin{eqnarray}
\ddot{R}_{1_B} + \kappa^2 R_{1_B} & = &
\frac{m_B}{\omega}\cos\omega t'  , \\
\ddot{R}_{1_S} + \kappa^2 R_{1_S} & = &
\frac{m_S}{\Delta\omega}\cos\Delta\omega t'' ,
\end{eqnarray}
of the form exactly like Wada's equation (4), the solution of which we already 
know. The steady-state solution for $R_1$ in a doubly barred potential
consists then of two terms and can be written in the form
\begin{equation}
R_1 =  -\frac{A_{0B}}{\omega} a_B \cos [2\varphi_{0s} + \omega (t-t_s)]
 -\frac{A_{0S}}{\Delta\omega} a_S \cos [2\varphi_{0s} + \omega (t-t_s) - \omega_p t] ,
\label{r1fin}
\end{equation}
where $A_{0B,0S}=1/p_{B,S}^2$, $a_{B,S}=p_{B,S}m_{B,S}$, 
$p_B=\omega^2-\kappa^2$, $p_S=\Delta\omega^2-\kappa^2$.
A similar solution can be written for $\varphi$. There
is an obvious way to extend this method to the arbitrary number of $N$ bars
by performing a separation of $N$ corresponding variables.

Clearly, the orbits in a potential of two independently rotating bars do 
not close. But from equation (\ref{r1fin}) we can see that the time 
transformation $t \rightarrow t+2\pi/\omega_p$ is equivalent to the initial
angle transformation 
$\varphi_{0s} \rightarrow \varphi_{0s} + \pi \omega / \omega_p$. 
This means
that after time $2\pi/\omega_p$, when the two bars return to the same 
relative orientation, the particle ends up at a place, which at the 
last time was occupied by another particle, with its guiding center
at the same radius, but position angle differing by 
$\pi \frac{\omega}{\omega_p}$.
Therefore particles having the same guiding radius interchange positions
at consecutive bar alignments, while remaining on the curve appropriate
to the guiding radius $R_0$. This
curve is the epicyclic approximation to {\it the loop}; in this linear 
approximation, a loop is a set of particles 
which have the same guiding radius, and lack any free epicyclic motion,
but respond only to the periodic forcing of the two bars.
A general orbit with that same guiding center will oscillate about the loop.
Loops in the epicyclic approximation are presented for Model 1 in Maciejewski 
\& Sparke (1997).

\subsection{The case of dissipative motion --- the Damped Epicyclic 
Approximation}
Lindblad \& Lindblad (1994) and Wada (1994) showed that if a dissipative
term is added to the equations of motion of a particle orbiting in a 
single rotating bar, the pattern of closed orbits in linearized solution 
closely reflects the gas streamlines in flows modeled
hydrodynamically. This {\it Damped Epicyclic Approximation} gives us a cheap
way to preview the hydrodynamic simulation. Obviously, this approach
cannot reproduce a shock structure, or give the inflow rate.

Adding a friction term $\bf F_{\rm fric}$ to equation (\ref{eqmot}) we get
\begin{equation}
\label{eqmotF}
\ddot{\bf r} = -\nabla \Phi  - 2 ({\bf \Omega_B} \times \dot{\bf r})
+ |{\bf \Omega_B}|^2  {\bf r} + {\bf F_{\rm fric}}.
\end{equation}
A true friction term ${\bf F_{\rm fric}}$ should be proportional to the 
velocity gradient in the gas, but this information is not available 
{\it a priori}, so we must use other forms of the dissipative term. 
We may postulate that the frictional force is proportional to the difference 
between the particle velocity $\dot{\bf r}$ and the circular velocity 
${\bf \Omega} \times {\bf r}$ at its position in an axisymmetric potential
$\Phi_0$, which is zero for a particle in circular orbit in that potential.
The dissipative force ${\bf F_{\rm fric}}$ in the 
inertial frame can be written as
\begin{equation}
\label{fric1a}
{\bf F_{\rm fric}} = - 2 \lambda ( \dot{\bf r}_i \; - 
\; {\bf \Omega} \times {\bf r}_i ) ,
\end{equation}
where ${\bf r}_i$ is the particle's position in the inertial frame, and 
$\lambda$ is the friction coefficient: the minus sign means
that dissipation tends to reduce velocity deviations for
positive $\lambda$. The friction described by Lindblad \& Lindblad (1994)
leads to this same expression for ${\bf F_{\rm fric}}$,
so we call this form the `Lindblad friction'.

The rotation speed $\dot{\bf r}_i$ in the inertial frame is related to the
speed $\dot{\bf r}$ in a frame rotating with the primary bar by
$\dot{\bf r}_i = \dot{\bf r} + {\bf \Omega_B} \times {\bf r}_i$,
thus equation (\ref{fric1a}) becomes
\begin{equation}
\label{fric1b}
{\bf F_{\rm fric}} = - 2 \lambda \left[ \dot{\bf r}
- ({\bf \Omega}-{\bf \Omega_B}) \times {\bf r} \right] .
\end{equation}
The bar pattern speed enters the friction formula
because the dissipative term depends on both radial and tangential velocity
deviations from the circular orbit. Wada (1994) used another form of 
friction, which depends on radial velocity deviations only:
\begin{equation}
\label{fric2}
{\bf F_{\rm fric}} =
- 2 \lambda ( \dot{\bf r} \cdot {\bf r}) \cdot {\bf r} / |{\bf r}|^2 .
\end{equation}
In dissipative motion, the two components of particle velocity are
coupled, which makes these two forms of friction compatible.

For Lindblad's friction, we get in a single bar the
second order equation for the radial correction $R_1$
\begin{equation}
\label{zetabisF}
\ddot{R_1} + 4\lambda\dot{R_1} + 4(\Omega_0^2+\lambda^2-\Omega_0A) R_1
= \frac{m}{\omega}\cos\omega t' + \frac{n}{\omega}\sin\omega t' ,
\end{equation}
where $t'$ is the same as defined for equation (\ref{dblezetabis}),
and $n=2\lambda C$. When we use Wada's friction defined in equation 
(\ref{fric2}), the second order equation for $R_1$ takes the form
\begin{equation}
\label{zetabis2F}
\ddot{R_1} + 2\lambda\dot{R_1} + 4\Omega_0 R_1 (\Omega_0-A)
= \frac{m}{\omega} \cos\omega t' .
\end{equation}
The epicyclic frequency $\kappa$ depends on the friction strength
for the Lindblad friction, but is the same as in the case with no 
friction when Wada's friction is used. The damping term is half as large 
for Wada's friction, which damps radial deviations only. 

For Lindblad's friction, the dissipative counterpart of the second order 
differential equation (\ref{dblezetabis}) for $R_1$ in the doubly barred 
case can be written as
\begin{equation}
\label{dblezetabisF}
\ddot{R_1} + 4\lambda\dot{R_1} + \kappa^2 R_1 =
\frac{m_B}{\omega}\cos\omega t' +\frac{n_B}{\omega}\sin\omega t' +
\frac{m_S}{\Delta\omega}\cos\Delta\omega t''
+\frac{n_S}{\Delta\omega}\sin\Delta\omega t'',
\end{equation}
where the additional coefficients are $n_B=2\lambda C_B$ and
$n_S=2\lambda C_S$. In presence of friction, we get the following 
steady-state solution for $R_1$ in a doubly barred potential
\begin{eqnarray}
R_1 =  &  -\frac{A_{0S}}{\Delta\omega} & \left( a_S \cos [2\varphi_{0s} + \omega (t-t_s) - \omega_p t]
                                     + b_S \sin [2\varphi_{0s} + \omega (t-t_s)
- \omega_p t]
\right) \nonumber \\
& -\frac{A_{0B}}{\omega} & \left( a_B \cos [2\varphi_{0s} + \omega (t-t_s)]
                                     + b_B \sin [2\varphi_{0s} + \omega (t-t_s)]
 \right) ,
\end{eqnarray}
where for a dissipative case $a_{B,S}=p_{B,S}m_{B,S}+q_{B,S}n_{B,S}$, 
$b_{B,S}=p_{B,S}n_{B,S}-q_{B,S}m_{B,S}$, $q_B=4\lambda\omega$, and
$q_S=4\lambda\Delta\omega$. A corresponding solution can be written 
for $\varphi$. Like in the frictionless
case, also in the damped epicyclic approximation particles with the same
guiding radius remain on the same loops. These loops for Model 1 are
presented in Maciejewski \& Sparke (1997).

\section{Potential of a prolate Ferrers' bar}
Here we derive the potential of a prolate Ferrers' bar by reducing 
Pfenniger's (1984) solution for a triaxial bar. It can be expressed by 
a third order polynomial in $x^2$, $y^2$, $z^2$ with coefficients 
$\omega_{im}$ (the reduced version of Pfenniger's coefficients $W_{ijk}$) 
defined by
\begin{equation}
\omega_{im} (x,y,z) = \int_{\lambda}^{+\infty} 
          \frac{du}{\sqrt{a^2+u}(b^2+u)}
          \frac{1}{(a^2+u)^i} \frac{1}{(b^2+u)^m}
\end{equation}
The definition of $\lambda$, and $\Delta(\lambda)$ below, can be found
in Pfenniger's paper (see also Binney \& Tremaine 1987, pp. 57 and 61). 
Since $\lambda=0$ inside the bar, $\omega$'s do not
depend on location there. Pfenniger's recurrence relations of (A11) type
reduce to 
\begin{equation}
\label{omegaim}
\omega_{im} = (\omega_{i-1,m} - \omega_{i,m-1}) / (a^2 - b^2) .
\end{equation}
with
\begin{eqnarray}
\omega_{i0} & = & \frac{2}{2i-1} [
\frac{1}{\Delta(\lambda)(a^2+\lambda)^{(i-1)}} - \omega_{i-1,1} ] , \\
\label{omega0m}
\omega_{0m} & = & \frac{1}{2m}   [
\frac{2}{\Delta(\lambda)(b^2+\lambda)^{(m-1)}} - \omega_{1,m-1} ] , \\
\omega_{00} & = &        \ln \left( \frac{1+\sin\theta}{\cos\theta} \right) 
                  \frac{2}{\sqrt{a^2-b^2}} , \\
\omega_{10} & = & \left[ \ln \left( \frac{1+\sin\theta}{\cos\theta} \right)  - \sin \theta \right]
                  \frac{2}{(a^2-b^2)^{3/2}} , \\
\omega_{01} & = & \frac{\tan^2\theta \sin\theta}{(a^2-b^2)^{3/2}} - \omega_{10}/2 ,
\end{eqnarray}
where $\cos\theta =  \sqrt{(b^2+\lambda)/(a^2+\lambda)}$.
Then the potential of a prolate Ferrers bar can be written as
\begin{eqnarray}
\Phi(x,y,z) & = & -\frac{\rho_0 ab^2}{3} \left( \omega_{00} - 6x^2y^2z^2 \omega_{12} \right. \nonumber \\
     &   & + x^2\{x^2[3\omega_{20} - x^2\omega_{30}] +
3[y^2(2\omega_{11}-y^2\omega_{12}-x^2\omega_{21}) - \omega_{10}]\} \nonumber \\
     &   & + y^2\{y^2[3\omega_{02} - y^2\omega_{03}] +
3[z^2(2\omega_{02}-z^2\omega_{03}-y^2\omega_{03}) - \omega_{01}]\} \nonumber \\
& & \left. + z^2\{z^2[3\omega_{02} - z^2\omega_{03}] +
3[x^2(2\omega_{11}-x^2\omega_{21}-z^2\omega_{12}) - \omega_{01}]\} \right)
\end{eqnarray}
The recipe above is equivalent to the formulae given by Perek (1962) and
de Vaucouleurs \& Freeman (1972).

\begin{thebibliography}{99}
\bibitem{a1} Athanassoula, E.  1992a  MNRAS 259, 328
\bibitem{a2} Athanassoula, E.  1992b  MNRAS 259, 345
\bibitem{bs} Benedict, G.F., Smith, B.J. \& Kenney, J.D. 1996 AJ 111, 1861
\bibitem{bt} Binney, J.  \& Tremaine, S. 1987 `Galactic Dynamics' (Princeton University Press)
\bibitem{bc} Buta, R. \& Crocker, D.A. 1993 AJ 105, 1344
\bibitem{c1} Combes, F. 1994 in `Mass Transfer Induced Activity in Galaxies', ed. I. Shlosman (Cambridge University Press), p170
\bibitem{cg} Contopoulos, G., \& Grosb{\o}l, P. 1989 A\&AR 1, 261 
\bibitem{cp} Contopoulos, G., \& Papayannopoulos, Th. 1980 A\&A 92, 33 
\bibitem{vf} de Vaucouleurs, G. \& Freeman, K.C. 1972 Vistas Astron. 14, 163
\bibitem{dk} Devereux, N.A., Kenney, J.D.P. \& Young, J.S. 1992 AJ 103, 784
\bibitem{es} Erwin, P. \& Sparke, L.S. 1999 ApJ 521, L37
\bibitem{f1} Friedli, D. 1996 in `Barred Galaxies', IAU Colloq 157, eds. R. Buta \etal\, ASP Conf. Ser., p378
\bibitem{fm} Friedli, D. \& Martinet, L. 1993 AAp 272, 27
\bibitem{fw} Friedli, D., Wozniak, H., Rieke, M., Martinet, L. \& Bratschi, P. 1996 AApS 118, 461
\bibitem{hs} Heller, C.H. \& Shlosman, I. 1994 ApJ 424, 84
\bibitem{jc} Jungwiert, B., Combes, F. \& Axon, D.J. 1997, A\&AS 125, 479
\bibitem{k1} Kuzmin, G. 1956 Astron. Zh. 33, 27
\bibitem{ll} Lichtenberg, A.J., \& Lieberman, M.A. 1992, Regular and Chaotic Dynamics, 2nd edition, (New York: Springer)
\bibitem{li} Lindblad, P.O.  \&  Lindblad, P.A.B. 1994 in `The Physics of Gaseous and Stellar Disks of Galaxies', ed. I.R. King, ASP Conf Ser. Vol 66, p29
\bibitem{lg} Louis, P.D. \& Gerhard, O.E. 1988 MNRAS 233, 337
\bibitem{m1} Maciejewski, W. 1999 in `Galaxy Dynamics: from the Early Universe to the Present', 15th IAP Meeting, Paris 9-13 July 1999
\bibitem{ms} Maciejewski, W. \& Sparke, L.S. 1997 ApJ 484, L117
\bibitem{mr} Mulchaey, J.S., Regan, M.W. \& Kundu, A. 1997 ApJS 110, 299
\bibitem{p1} Perek, L. 1962 Adv. Astron. Astroph. 1, 65
\bibitem{p2} Pfenniger, D. 1984 AAp 134, 373
\bibitem{pn} Pfenniger, D. \& Norman, C. 1990 ApJ  363, 391
\bibitem{sw} Sellwood, J.A. \& Wilkinson, A. 1993 Rep. Prog. Phys. 56, 173
\bibitem{sa} Shaw, M.A., Axon, D., Probst, R. \& Gatley, I. 1995 MNRAS 274, 369
\bibitem{ts} Tagger, M., Sygnet, J.F., Athanassoula, E., \& Pellat, R. 1987 ApJ 318, L43
\bibitem{t1} Toomre, A. 1963 ApJ 138, 385
\bibitem{w1} Wada, K. 1994 PASJ 46, 165
\bibitem{w2} Wozniak, H. 1991 Ph.D. Thesis
\bibitem{wf} Wozniak, H., Friedli, D., Martinet L., Martin, P. \& Bratschi, P. 1995 ApJ Supp 111, 115
\end{thebibliography}
\end{document}